\documentclass[english,aps,prd,onecolumn,
superscriptaddress,preprintnumbers,nofootinbib]{revtex4-2}
\usepackage{geometry}
\usepackage{subcaption}  
\usepackage{graphicx}
\usepackage[utf8]{inputenc}
\usepackage{amssymb}
\usepackage{verbatim}
\usepackage{amsmath}
\usepackage{color}
\usepackage{hyperref}
\usepackage{fancyhdr}
\usepackage{graphicx,color}
\hypersetup{colorlinks,linkcolor={black},citecolor={blue},urlcolor={blue}}

\newcommand{\be}{\begin{equation}}
\newcommand{\ee}{\end{equation}}
\newcommand{\bea}{\begin{eqnarray}}
\newcommand{\eea}{\end{eqnarray}}
\newcommand{\pa}{\partial}
\newcommand{\eps}{\epsilon}

\newcommand{\morder}[1]{{\cal O}\left(#1 \right)}

\usepackage{xcolor}

\definecolor{ogreen} {RGB}{0,0,0}
 
\definecolor{oblue} {RGB}{91,125,191}
 
\definecolor{ored} {RGB}{255,0,0}

\newcommand{\rl}[1]{\textcolor{ogreen}{#1}}

\begin{document}
\title{Holographic transport in anisotropic plasmas}
\author{Tuna Demircik}
\affiliation{Institute for Theoretical Physics, Wroclaw University of Science and Technology, 50-370 Wroclaw,
Poland}
\author{Domingo Gallegos}
\affiliation{Facultad de Ciencias, Universidad Nacional Aut\'onoma de M\'exico, Investigaci\'on Cient\'ifica C.U., 04510 Coyoacan, Ciudad de Mexico, Mexico }
\author{Umut G\"ursoy}
\affiliation{Institute for Theoretical Physics and Center for Extreme Matter and Emergent Phenomena, Utrecht University, Leuvenlaan 4, 3584 CE Utrecht, The Netherlands}
\author{Matti J\"arvinen}
\affiliation{Asia Pacific Center for Theoretical Physics, Pohang, 37673, Korea}
\affiliation{Department of Physics, Pohang University of Science and Technology, Pohang, 37673, Korea}
\author{Ruben Lier}
\affiliation{Institute for Theoretical Physics, University of Amsterdam, 1090 GL Amsterdam, The Netherlands}
\affiliation{Dutch Institute for Emergent Phenomena (DIEP), University of Amsterdam, 1090 GL Amsterdam, The Netherlands}


\begin{abstract}
We study energy-momentum and charge transport in strongly interacting holographic quantum field theories in an anisotropic thermal state by contrasting three different holographic methods to compute transport coefficients: standard holographic calculation of retarded Green's functions, a method based on the null-focusing equation near horizon and the novel method based on background variations. Employing these methods we compute anisotropic shear and bulk viscosities and conductivities with anisotropy induced externally, for example by an external magnetic field. We show that all three methods yield consistent results. The novel method allows us to read off the transport coefficients from the horizon data and express them in analytic form from which we derive universal relations among them. Furthermore we extend the method based on the null-focusing equation to Gauss-Bonnet theory to compute higher derivative corrections to the aforementioned transport coefficients.
\end{abstract}
\preprint{APCTP Pre2024 - 007}
\maketitle
\tableofcontents
\pagebreak

\section{Introduction}

Quark-gluon plasma produced in heavy-ion collisions is well described as a strongly interacting relativistic fluid, see \cite{Heinz:2013th,Teaney:2000cw,Romatschke:2007mq,Busza:2018rrf} for reviews. Anisotropies in pressure gradients in directions transverse to the beam, that are present in off-central collisions, played an important role in establishing this hydrodynamic description early on, through matching the experimentally observed flow parameters to hydrodynamic simulations, see \cite{JETSCAPE:2020mzn,Nijs:2020roc,Nijs:2020ors} for recent updates. There is a second source of anisotropy that arises from expansion of the plasma along the beam direction which is faster than its transverse expansion. Finally, intense magnetic fields\footnote{Large vorticities are
produced in
off-central collisions which
break the rotational invariance in the plasma \cite{STAR:2018gyt} also leading to anisotropic transport.  We will however not discuss vortical effects in this paper.} are produced in off-central collisions, leading to yet another source of anisotropy in the plasma. Indeed, a back-of-the-envelope estimate based on the Biot-Savart law 
leads to a maximal magnetic field of $B \approx 10^{15}-10^{19} $ G at RHIC and $B \approx  10^{16}-10^{21}$ G at LHC~\cite{Kharzeev:2007jp,Skokov:2009qp,Tuchin:2013ie,Voronyuk:2011jd,Deng:2012pc,Tuchin:2010vs,McLerran:2013hla,Gursoy:2014aka}.

Another important physical system subject to extremely high magnetic fields is a neutron star. The strength of the magnetic field can reach $10^{15}$ G on the surface of a magnetar~\cite{Haensel:2007yy} and $10^{16}$ G inside it \cite{Makishima:2014dua,DallOsso:2018dos}. 
Moreover, the strength can be amplified during a binary merger and can exceed $10^{16-17}$ G \cite{Price:2006fi,Giacomazzo:2014qba,Kiuchi:2015sga,Kiuchi:2017zzg}, overlapping with the values of the magnetic field produced in heavy-ion collisions \cite{Lovato:2022vgq} in the later times of the plasma evolution before freezout. Although it is still an open question whether neutron stars with quark matter cores exist, the characteristics of neutron star matter have been demonstrated to agree better with the anticipated features of nearly conformal quark matter than with those of nuclear matter \cite{Annala:2019puf,Annala:2023cwx}. 
Even if quark matter does not exist inside isolated neutron stars, it can be created in neutron star binary mergers~\cite{Most:2018eaw,Bauswein:2018bma,Tootle:2022pvd}. When the anticipated extremely large magnetic fields during a merger event are taken into account, the deconfined quark matter accommodates magnetic field-induced anisotropy. As suggested in~\cite{Alford:2017rxf}, neutron star matter during a merger event is not only sensitive to the equation of state but also to the 
bulk viscosity. A complete understanding of the transport properties requires the inclusion of effects arising from anisotropy, even though these effects might be subdominant.    

Despite abundant experimental evidence and numerical hydrodynamic simulations, transport of energy-momentum and charge in these anisotropic plasma states has not been adequately addressed from a theoretical point of view. There are two obvious reasons for this: First is the insufficiency of perturbative QCD in computing associated transport coefficients in the presence of strong correlations. The second is the shortfall of lattice QCD methods, which are suitable to compute Euclidean rather than retarded correlation functions which are required in Kubo formulas for transport. These shortcomings hold as well in the isotropic plasma states. Nevertheless, while further advances in computing isotropic transport coefficients in lattice QCD are within reach by analytic continuation from Euclidean to real-time correlators, see e.g. \cite{Meyer:2011gj}, this will be much harder to achieve in anisotropic states. 

Gauge-gravity duality \cite{Maldacena:1997re,Gubser:1998bc,Witten:1998qj} comes into play here. Anisotropic plasma states in holographic quantum field theories have been extensively studied in applications to condensed matter, see e.g. \cite{Zaanen:2015oix,Hartnoll:2016apf} for recent reviews, but also in applications to QCD and the quark-gluon plasma in \cite{Azeyanagi:2009pr, Mateos:2011tv,Mateos:2011ix,Giataganas:2012zy,Chernicoff:2012iq,Giataganas:2013hwa,Rougemont:2014efa,Rougemont:2015oea,Drwenski:2015sha,Fuini:2015hba,Arefeva:2016phb,Gursoy:2016tgf,Demircik:2016nhr,Gursoy:2016ofp,Gursoy:2018ydr,Gursoy:2020kjd}. On the gravitational side these states are realized by planar blackhole solutions with coordinates $(r, t, x_i)$ where $r$ is the holographic direction, $t$ is time and $x_i$ are the spatial field theory directions, with the spatial metric $g_{ij}$ being diagonal with unequal components. These holographic studies include plasmas where the anisotropy is due to an external magnetic field, see \cite{Kharzeev:2013jha,Gursoy:2016ebw,Gursoy:2021kqt} for reviews. 

The main purpose of our paper is to explore transport of conserved quantities, energy-momentum and charge in anisotropic plasma states in strongly coupled holographic quantum field theories. We will do this using various different approaches: (i) traditional holographic technique to compute retarded Green's functions from fluctuations of the background; we will refer to this as the Gubser-Pufu-Rocha (GPR) method after \cite{Gubser:2008sz} which improved the standard holographic computation and applied it to computation of the bulk viscosity, (ii) extending the method of  \cite{Eling:2011ms} --- based on the null-focusing equation near the horizon --- to anisotropic black holes which we refer to as the Eling-Oz (EO) method, and, (iii) using a novel method \cite{Demircik:2023lsn} that relates transport coefficients directly to horizon data by varying parameters of the background. We refer to this latter method as ``background variation'' (BV) for short. 

An obvious motivation to invoke the gauge-gravity duality here is that the strong coupling limits of transport coefficients, such as the shear viscosity to entropy ratio obtained from the duality \cite{Policastro:2001yc} are much closer to observation \cite{JETSCAPE:2020mzn,Nijs:2020roc,Nijs:2020ors} than perturbative QCD results \cite{Arnold:2000dr,Arnold:2003zc}. In addition to shear, in non-conformal but isotropic plasmas, energy-momentum transport is characterized also by the bulk viscosity which has also been extensively studied in holography \cite{Parnachev:2005hh,Benincasa:2005iv,Buchel:2005cv,Buchel:2007mf,Gubser:2008sz,Buchel:2008uu,Gursoy:2009kk,Gursoy:2010jh,Buchel:2011wx,Buchel:2011uj,Eling:2011ms,Buchel:2011yv,Hoyos:2020hmq}. There is however an important difference between the holographic computation of shear and bulk viscosity. Shear viscosity is obtained from the transversely polarized graviton fluctuation on AdS black-hole and does not mix with the other fluctuations, leading to a single massless fluctuation equation. \rl{For isotropic and homogeneous fluids}, the result is universal, that is, independent of presence and details of the other sectors in the gravitational action e.g. scalar potentials, and given by $\eta/s = 1/4\pi$. It is easy to understand the reason for this universality. As clearly explained in \cite{Gubser:2008sz}, in the absence of a mass term in the fluctuation equation, one can define a ``graviton flux'' that is conserved, hence its value on the boundary and at the horizon becomes the same. Thus one can read off the transport coefficient directly from the horizon which leads to universal values. The same holds for electric conductivity which can also be directly read off from the horizon, see \cite{Iqbal:2008by}\footnote{ The horizon formula for conductivity leads to ``bulk universality" and it does not imply any universality from the boundary perspective. However, it has important applications in quantum critical phenomena \cite{Kovtun:2008kx,Herzog:2007ij}. }.

This argument is also  valid for shear viscosity in non-conformal holographic theories\footnote{The result is modified if one goes beyond the two-derivative (super)gravity approximation \cite{Brigante:2007nu}, see \cite{Cremonini:2011iq} for a review and \cite{Buchel:2023fst} for a recent study. Adding such corrections is important to reproduce the temperature dependence of $\eta/s$ observed in heavy-ion collisions \cite{JETSCAPE:2020mzn,G_rsoy_2016,Nijs:2020ors} as was shown in \cite{Cremonini:2012ny}.} as presence of other matter fields that break conformal invariance of the dual field theory does not affect universality at the horizon \cite{Buchel:2003tz}. 

This ceases to be the case for the ``massive'' fluctuations, e.g. dual to bulk viscosity, which maps to trace of the graviton fluctuation and mixes with fluctuation of other scalars in the dual background. Even though one can again construct a conserved flux \cite{Gubser:2008sz,Kaminski:2009dh}, its value depends on a non-trivial coefficient that can only be fixed by determining the full solution to the fluctuation equation from boundary to horizon which is now hard because of the mass term. The bottomline is that horizon data itself does not seem to be sufficient to fix transport coefficients that are dual to massive fluctuations.  

An independent approach was put forward by Eling and Oz in \cite{Eling:2011ms}, see also \cite{Eling:2010hu,Eling:2009sj} based on relating the null-focusing (or Raychaudhuri) equation \cite{Raychaudhuri:1953yv} near the horizon to positivity of entropy production in the dual QFT. This approach does determine the transport coefficients in terms of the horizon data, however it is build on positive entropy production. In particular, it will not work for non-dissipative transport, e.g. anomalous conductivities,  which does not contribute to entropy production (see e.g. \cite{Landsteiner:2011cp,Landsteiner:2012kd,Landsteiner:2016led}). Certain type of coefficients associated to dissipationless transport, such as anomalous conductivities in the presence of `t Hooft anomalies, turn out to be free of radiative corrections hence can be obtained directly in perturbative QFT without resorting to holography. However, contribution of dynamical gauge fields to anomalous transport at strong coupling remains an open problem. Even though this has been addressed in holography in \cite{Gursoy:2014ela,Jimenez-Alba:2014iia,Gallegos:2018ozs}, analytic expressions that relate transport to horizon data are hard to get\footnote{Analytic or semi-analytic expressions are essential to obtain generic relations between transport coefficients, and hence highly desirable}. Nevertheless, the approach of Eling and Oz (EO) will be crucial to our paper. We will not only prove the EO formula for bulk viscosity using the traditional holographic techniques, but also extend them to anisotropic states and to include a certain type of higher derivative corrections, in particular in the Gauss-Bonnet theory. 

Turning back to the our main focus of anisotropic transport, we notice that computation of these ``massive'' transport coefficients in {\em anisotropic states} in the traditional approach will be quite cumbersome. First, the number of transport coefficients multiply in the presence of anisotropies.  Now there will be two shear viscosities: one when shear deformation is on the plane transverse to the direction of anisotropy\footnote{We will assume a single anisotropic direction in this paper.}, and another when it's along the direction of anisotropy. Similarly, there are two electric conductivities and three bulk viscosities, see section~\ref{sec:magneticviscosity}. Second, the bulk fluctuations dual to a generic fluctuation will mix other modes becoming ``massive''. 

These difficulties led us in our companion paper \cite{Demircik:2023lsn}, to introduce the BV method which computes generic transport coefficients directly from the horizon data. This is based on the realization that, as the transport coefficients in QFT are computed from vanishing frequency limit of retarded Green's functions, they can be naturally mapped to variations of the background parameters, such as charge and mass, in the dual gravitational blackhole background. We introduced this method in \cite{Demircik:2023lsn}, which we then used to calculate anisotropic bulk viscosities. We also compared with GPR and EO using the holographic QCD model of \cite{Jarvinen:2011qe,Gursoy:2007er,Gursoy:2007cb} --- see \cite{Jarvinen:2021jbd} for a review --- finding perfect agreement.  

In the present work, we provide derivation of the BV method in more detail and expand on its applications to anisotropic transport coefficients, including anisotropic conductivities and charged plasmas. Moreover we extend the EO method to include anisotropies and higher derivative corrections which allows for computing small `t Hooft coupling corrections to transport coefficients. 

The structure of this work is as follows. In Sec.~\ref{sec:istropicviscosity}, we review the computation of bulk viscosity by Gubser, Pufu and Rocha \cite{Gubser:2008sz}, which requires solving fluctuation equations. We then show how this result can be incorporated into background variations. In Sec.~\ref{sec:magneticviscosity} we use this method to obtain an expression for the magnetic bulk viscosities in terms of background variations. In Sec.~\ref{sec:magneticviscosity}, we also discuss magnetic conductivity and a universal relation between magnetic conductivity and magnetic shear viscosity. In Sec.~\ref{sec:EO}, We show how the obtained result coincides with an alternative method for computing transport at the horizon, which was first used by Eling and Oz \cite{Eling:2011ms}. In addition, we investigate here anisotropic transport in a class of higher derivative theories given by the Gauss-Bonnet correction.  In Sec.~\ref{sec:numericalcheck}, we employ a specific holographic model~\cite{Jarvinen:2011qe,Gursoy:2007er,Gursoy:2007cb} (with the full back reaction of flavor sector) to obtain numerical results for the viscosities based on the derived expressions. Finally, we conclude with the discussion in the Sec. ~\ref{sec:disc}. 

\section{Isotropic viscosity}
\label{sec:istropicviscosity}
\subsection{Viscosity with GPR method}
\label{sec:gubsergublk}
As explained in the Introduction, we will switch back and forth between the three different holographic methods to compute transport coefficients. We first review the holographic bulk viscosity computation put forward by Gubser, Pufu and Rocha in Ref.~\cite{Gubser:2008sz} and generalized and further developed by \cite{Kaminski:2009dh} and others. After reviewing the GPR method in the case of isotropic states below, we will then extend this method to study transport in anisotropic states and, in addition, use this as a starting point to introduce a novel means to compute transport coefficients that we coin ``background variations'' \cite{Demircik:2023lsn}. 

For an isotropic and relativistic fluid conservation laws for energy-momentum and charge are given by
\begin{align}
    \partial_{\mu} T^{\mu \nu } = F^{\mu \nu } J_{\nu} ~~ , ~~  \partial_{\mu} J^{\mu} =0  ~~ . 
\end{align}
Given the four-velocity $u_{\mu}$ of the fluid, the constitutive equations can be written as
\begin{align}\label{someCurrents}
J^{\mu} = q u^{\mu} + J_{(1)}^{\mu } ~~ , ~~     T^{\mu \nu }  = \epsilon u^{\mu} u^{\nu} +  p \Delta^{\mu \nu }  + T_{(1)}^{\mu \nu} ~~,  ~~     \Delta^{\mu \nu }   =  \eta^{\mu \nu } + u^{\mu } u^{\nu} ~~ ,  
\end{align}
where 
the quantities with subscript (1) denote the dissipative contributions at first derivative order. 
At this order $T_{(1)}^{\mu \nu} $ is given by 
\begin{align}  \label{eq:stre9897887ss}
\begin{split}
    T_{(1)}^{\mu \nu}  = &  - 2 \eta \hat{\Delta}^{\mu \nu \alpha \beta } \partial_{\alpha} u_{\beta }    -  \zeta \Delta^{\mu \nu } \Delta^{ \alpha \beta } \partial_{\alpha} u_{\beta }  ~~  ,  ~~ \hat{\Delta}^{\mu \nu \alpha \beta }  =   \Delta^{\mu ( \alpha } \Delta^{ \beta )\nu} -  \frac{1}{3} \Delta^{\mu \nu } \Delta^{ \alpha \beta }  ~~ ,  
\end{split}
  \end{align}whereas for $J_{(1)}^{\mu}$ we have $J_{(1)}^{\mu} =  - \sigma  T \Delta^{\mu \nu } \partial_{\nu} \frac{\mu}{T} $. Here we denote $A^{\cdots (\mu}B^{\nu)\cdots} = (A^{\cdots \mu}B^{\nu\cdots}+A^{\cdots \nu}B^{\mu\cdots})/2$.
  In holographic computations, the shear viscosity was found to be universally given by \cite{Kovtun:2004de}
\begin{align}
    \eta = \frac{s}{4 \pi } ~~~ . 
\end{align}
For a conformal fluid, the bulk viscosity is zero, so to obtain a non-vanishing bulk viscosity we must consider a fluid that is non-conformal, which can be done by introducing a dilaton potential in the dual gravity action. We then consider the Einstein-dilaton theory
\be\label{eq:Sdef}
 S = \mathcal{N} \int d^5 x \sqrt{-g}\left[R -\frac{1}{2}
 (\pa \phi)^2 - V(\phi) +\mathcal{L}_\mathrm{matter}(\phi,F) \right]  = \mathcal{N} \int d^5 x ~\mathcal{L}~~,  
\ee
\rl{where the precise form of the matter action is not important. We can take it to be the Maxwell Lagrangian as we will do below. }
We consider the following metric ansatz
\be \label{eq:metric}
 ds^2 = e^{2A(r)}\left(f(r)^{-1}dr^2 -f(r)dt^2 + d\mathbf{x}^2  + 
  H_{MN}(r,t) dx^M dx^N \right) \ ,  
\ee
where $H_{MN}(r,t)$ are fluctuations. We can use diffeomorphism symmetry to choose a gauge where the dilaton fluctuation is set to zero \cite{Gubser:2008sz}. To compute bulk viscosity, we only need to turn on the components related to the scalar sector, which are $H_{tt}$, $H_{rr}$, $H_{rt}$, as well as the spatial trace, whereas we turn off dilaton fluctuations. To be specific, we write 
\begin{align}
\begin{aligned}
 H_{11} &= H_{22} = H_{33} = h(r)e^{-i\omega t} \\ 
  H_{tt} &= - f(r) (\Theta(r) + h(r))e^{-i\omega t} 
 \end{aligned}
 &&
 \begin{aligned}
      H_{rt} &= h_{rt}(r)e^{-i\omega t} \\
 H_{rr} &= f(r)^{-1}(\Gamma(r) + h(r))e^{-i\omega t}
  \end{aligned}
\end{align}
After eliminating the other fluctuations, we obtain a decoupled equation for $h$ which reads 
\be \label{hfluct}
 h^{\prime \prime} (r) + \left[ \frac{f'(r)}{f(r)} + \mathcal{K}_1  (r)\right]h'(r) + \left[\frac{\omega^2}{f(r)^2}+\mathcal{K}_2 (r )\right]h(r) = 0
\ee
where the background-dependent function \rl{$\mathcal{K}_1  (r) = 3A'+\frac{d}{dr}\log (\phi'^2/A'^2)$, while $\mathcal{K}_2  (r)$ depends on the choice of the matter action. It has a simple pole at the horizon} where $f(r)$ vanishes. Having derived the equation for $h$, one then writes down the bilinear Lagrangian which will enable one to obtain the Kubo formula for the bulk viscosity. 

We can then expand the Lagrangian in~\eqref{eq:Sdef} quadratically in the sources to find
\begin{align}
    \mathcal{L}^{(2)} = \partial_r J +  \vec{h}^{* T} \left( \frac{\partial \mathcal{L}}{ \partial \vec{h}^* } - \partial_r \frac{\partial \mathcal{L}}{ \partial  \vec{h}^{* \prime }   }  \right) ~~  , ~~  \vec{h} = \{ h , \Theta  , \Gamma     \}   , 
\end{align}
where the $J$-term is the only part that is non-vanishing on-shell. This term can be related to the retarded Green's function as \cite{Herzog_2003,Son_2002}
\begin{align}
\text{Im }  G^R (\omega )  = - \mathcal{N}  \mathcal{F}~~  ,  ~~ \mathcal{F} = - \text{Im }   J    ~~ , 
\end{align}
where the trace-trace retarded correlator in the QFT is given by 
\begin{align} \label{eq:greensfunctionrelation}
     G^R (\omega )   =  - i \int dt d^3x e^{i \omega t } \left\langle \left[ \frac{1}{2} T_{i}^{\, \, i } (t , \vec{x}) 
 ,   \frac{1}{2}  T_{k}^{\, \, k } (0,0 )  \right] \right\rangle ~~ . 
\end{align}
Here $\mathcal{F}$ is a conserved flux of background fluctuations which, therefore, can be computed at any $r$, and most easily near the horizon, where it is given by
    \begin{align}
 \label{eq:Fterm1}
\mathcal{F} = - i \lim_{r \rightarrow r_h } \exp(3 A ) \frac{ \phi^{\prime 2  } f  }{ 8   A'^2} \left( h' h^{*} - h^{* \prime } h \right)    ~~ . 
    \end{align}
   Near the horizon, Eq.~\eqref{hfluct} turns into
\begin{align} \label{eq:nearhorizondifferentialequation}
    h^{\prime \prime }    =  
  \frac{h^{\prime }   }{ r_h - r   }  - \frac{\omega^2  h }{f^{\prime 2}_h ( r_h - r )^2  }   + \mathcal{O} \left(r - r_h  \right)        ~~ . 
\end{align}
Solving Eq.~\eqref{eq:nearhorizondifferentialequation} and imposing infalling boundary conditions \cite{Son_2002} yields
\begin{align} \label{eq:horizonsolution1}
    h  =   c_-(\omega) (r_h - r )^{-  \frac{i \omega}{4 \pi T }}  =  c_-(\omega) \left[    1 -  \frac{i \omega}{4 \pi T }  \log (r_h- r)    \right]     ~~ ,  
\end{align}
where we used that $  T =  - \frac{1}{4 \pi } f^{\prime}_h$ and $c_-(\omega)$ is a function that in general depends on frequency $\omega$, however we will only need $c_- \equiv    c_-(0)$. Then, plugging Eq.~\eqref{eq:horizonsolution1} into  Eq.~\eqref{eq:Fterm1}, we find
\begin{align} \label{eq:GPRbulkviscosity}
    \zeta  =  -  \frac{4}{9}    \lim_{\omega \rightarrow 
  0  } \frac{1}{\omega} \text{Im } G^R (\omega )      = \frac{  s c^2_{-}   \phi^{\prime 2  }_h}{36 \pi  A_h'^2 } ~~ ,  
\end{align}
where $s$ is the entropy density given by 
\begin{align}
s =  4 \pi   \mathcal{N}  \exp(3 A_h)  ~~ ,     \end{align}
$A_h$ is the value of the scale factor $A$ at the horizon and $c_-$ can be obtained numerically by solving the equation for $h$ for $\omega =0 $ with the UV boundary condition  
\begin{align} \label{eq:UVboundarycondition}
    \lim_{r \rightarrow  r_b } h(r)  = 1 ~~ , 
\end{align}
where $r_b$ is the location of the boundary.
Clearly, the result (\ref{eq:GPRbulkviscosity}) depends on a quantity, i.e. $c_-$, that requires the full solution of the fluctuation equation from the horizon to the boundary. This can only be done numerically. Below we introduce a method which will allow one to express the result completely in terms of horizon data in a semi-analytic fashion.  

\subsection{Transforming fluctuations to background variations: case of zero magnetic field}
\label{sec:GPREOconnection}

In this section we show that the result of Eq.~\eqref{eq:GPRbulkviscosity}, which depends on a numerical constant that requires the full solution of the fluctuation equations, can instead be written in terms of variations of background fields. The idea is to relate the fluctuations in the limit of vanishing frequency, which one needs to obtain the Kubo formulas that give the transport coefficients, to variations of the background with respect to its parameters such as charge and temperature (or equivalently entropy). 

To proceed we first consider diffeomorphism gauge symmetries of the sources in a generic background. In the GPR computation, the gauge symmetries are used to turn off the dilaton fluctuation, which we will denote below as $ \chi (r)$. We will require the same. Under an infinitesimal diffeomorphism $r\mapsto r+\xi (r)$ the fluctuations transform as
\begin{align} 
    \label{eq:diffeo}
    \begin{aligned}
 \chi(r) &\to \chi(r) + \phi'(r) \xi(r)\,, \\       
  \Theta(r) &\to \Theta(r) + \frac{f'(r)}{f(r)}\xi(r) \,,
    \end{aligned}
    \begin{aligned}
        \qquad h(r) &\to h(r) +2 A'(r) \xi(r) \,,\\
 \Gamma(r) &\to \Gamma(r) - \frac{f'(r)}{f(r)}\xi(r) +2 \xi'(r)    \, .      \end{aligned}
 \end{align}
In order to eliminate $\chi(r)$, we take $\xi(r) = -\delta \phi(r)/\phi'(r)$. We can then write the fluctuations in Eq.~\eqref{eq:metric} as background variations,
\begin{align}
\begin{aligned}  
 \chi(r) &= 0  \,   \\ 
 \Theta(r) &=\frac{\delta f(r)}{f(r)} - \frac{f'(r)}{f(r)\phi'(r)}\, \delta \phi(r) \, \\ 
  & 
 \end{aligned}
 && 
 \begin{aligned}
h(r) &= 2 \delta A(r) - 2 \frac{A'(r)}{\phi'(r)}\, \delta \phi(r) \,,  \\ 
\Gamma(r) &= -\frac{\delta f(r)}{f(r)} + \frac{f'(r)}{f(r)\phi'(r)}\, \delta \phi(r) \\ 
  &   - 2 \frac{\delta\phi'(r)}{\phi'(r)}+ 2 \frac{\phi''(r)\delta\phi(r)}{\phi'(r)^2}
 \end{aligned}
 \label{mapping2} 
\end{align}
\rl{The mapping outlined above is not one-to-one because setting $\omega=0$ and taking $\omega\to 0$ limit do not necessarily coincide. To make these two equivalent, in particular,  one needs to start with the fluctuation equation and impose an additional constraint that arises from the $rt$-component of Einstein's equations. The $\omega\to 0$ limit of this fluctuation that satisfies the constraint would then coincide with the background variation defined at strictly $\omega =0$. 
}

\rl{That is, solving the equations~\eqref{mapping2} for the background variations, we obtain } 
\begin{align}
\label{solvedflucts}
\begin{aligned}
\frac{\delta \phi}{\phi'} &= -\frac12 \int_{r_b}^r d\tilde r \left(\Theta(\tilde r) + \Gamma(\tilde r)\right) + C_\phi\,,  \\
\delta A &= \frac12 h + \frac{A'}{\phi'}\delta \phi\,,\\
\delta f &= f \Theta + \frac{f'}{\phi'} \delta \phi\, ,
\end{aligned}
\end{align}
\rl{which is guaranteed to be a proper variation of the background so long as $h$, $\Theta$, and $\Gamma$ satisfy the full fluctuation equation, including the constraint coming from the $rt$-component of the Einstein equations\footnote{One might be worried that this argument fails because~\eqref{mapping2} is not an algebraic equation for the full variation, so that the solution contains the integration constant $C_\phi$. However, as we shall see explicitly below, this constant is identified with the infinitesimal constant variation of the $r$-coordinate (denoted by $\epsilon_r$ below), which confirms that the solution is a variation of a regular background.}.}

However, these relations do not completely absorb fluctuations into background variations yet, as the fluctuations are turned on at the UV boundary, whereas the variations of the background fields are not \rl{if the sources are kept fixed when the background is varied}. To turn off the fluctuations at the boundary as well, we further consider reparametrization symmetries of the background, which are given by 
\begin{subequations}
    \begin{align}
 r &\to r +r_0  ~~ ,  &&&\\
 A &\to A + \log \Lambda \,,&\qquad r &\to \Lambda r \,,&\qquad \rho &\to \Lambda^3 \rho  ~~ , &\\ 
 f &\to f_0 f\,,&\qquad A &\to A + \frac{1}{2}\log f_0\,,&\qquad \rho &\to f_0^{3/2} \rho  ~~ ,  &
\end{align}
\end{subequations}
 where $r$, $A$ and $f$ are the holographic coordinate, scale factor and the blackening factor in the conformal coordinate system (\ref{eq:metric}) with $H_{MN}$ set to zero, and $\rho$ is the charge density. These symmetries were shown to hold in the absence of charge in \cite{Gursoy:2008bu} and above is a natural generalization with charge. The infinitesimal versions of these symmetry parameters are 
\be
r_0 = \eps_r\,,\qquad  \Lambda = 1+\eps_\Lambda \,,\qquad f_0 = 1+\eps_f\,,
\ee
and they generate the following variation of the background:
   \begin{align}
       \label{symmvars1}
  \begin{aligned}
\delta_\epsilon A &= \epsilon_\Lambda+ (r \epsilon_\Lambda +\epsilon_r) A' +\frac{\epsilon_f}{2}  \\ 
\delta_\epsilon \phi &=  (r \epsilon_\Lambda +\epsilon_r) \phi'  
  \end{aligned}
  &&
  \begin{aligned}
 \delta_\epsilon f &=  (r \epsilon_\Lambda +\epsilon_r)f' + \epsilon_f f \\        \delta_\epsilon \rho &= (3 \eps_\Lambda + 3 \eps_f/2) \rho 
  \end{aligned}
 \end{align}
We can now use these infinitesimal reparametrizations to turn off the fluctuations at the boundary, i.e. we will undo the UV boundary condition of Eq.~\eqref{eq:UVboundarycondition}. The symmetry transformation of Eq.~\eqref{symmvars1} also modifies the UV value of $\Theta$. However, because only the $r$-derivative of $\Theta$ turns out to be physical, the UV value is arbitrary and we can take
\begin{align} \label{eq:UVboundaryconditiontheta}
    \lim_{r \rightarrow  r_b } \Theta (r)  = C_{\Theta} ~~ ,  
\end{align}
where $C_{\Theta}$ is an arbitrary constant that will drop out at the end of our computation. To turn off these UV values, we take the infinitesimal parameters to be
\be \label{epssol}
 \eps_f = - C_{\Theta}  \,,\qquad \eps_\Lambda = \frac{1}{2} (C_{\Theta} -1  ) ~~ .
\ee
\rl{Absence of the sources also sets $\epsilon_r = -C_\phi-r_b\eps_\Lambda$, but as it turns out, this parameter cancels in our analysis so its value is not important.}
Note that the values of these infinitesimal parameters are not necessarily small, this is due to the fact that the wave functions of the fluctuations, contrary to what their name suggests, are similarly not infinitesimal at the boundary (see~\eqref{eq:UVboundarycondition}). 

To obtain a formula for viscosity in terms of background fields we want to express the GPR expression for the bulk viscosity in (\ref{eq:GPRbulkviscosity}) in terms of background variations. Consider then the near horizon expansion of the background fields
\begin{subequations}
\begin{align}
 A(r) &= A_h + A_h'(r-r_h) + \morder{(r-r_h)^2}\,,&\quad f(r)& = f'_h(r-r_h) +\morder{(r-r_h)^2} & \,.
\end{align}     
\end{subequations}
Let us also write down the horizon expansion of the fluctuation equations
\begin{subequations}  \label{eq:expansions1}
\begin{align}
 h (r)  & =  h_h +  h'_h(r-r_h) + \morder{(r-r_h)^2} \,, \\
 \Theta  (r)  & = \frac{\Theta_r}{r - r_h } + \Theta_h + \Theta_h^{ I \prime}  (r-r_h) + \morder{(r-r_h)^2} \,.    
\end{align}
 \end{subequations}
\rl{ In addition, we note that $\Theta+\Gamma$ is regular at the horizon.}
 
 A crucial observation here is that the horizon value of $h$, i.e. $h_h$, becomes the same as the quantity that was introduced in (\ref{eq:horizonsolution1}) in the $\omega\to 0$ limit: $h_h = \lim_{\omega\to 0} c_-$. Moreover $\Theta_r $ is not independent but can be obtained from the fluctuation equations. Using the $rr$-component of the Einstein fluctuation equations near the horizon, we obtain
 \begin{align} \label{eq:nearhorizontheta}
    \Theta_r   = \frac{ h_h }{2  A_h^{\prime} }    ~~   , 
\end{align}
\begin{subequations}
so that for the total variation of the background fields, i.e. the sum of the variations in~\eqref{solvedflucts} and in~\eqref{symmvars1} \rl{ at the horizon}, we find
\begin{align}
 \delta\hat A &= \frac{1}{2}h_h -A_h' \delta r -\frac{1}{2}+\frac{A_h'}{2}r_h (C_{\Theta} - 1 )+\morder{r-r_h}\,,&\\
 \delta \hat f &= \frac{f_h'}{2 A_h'}h_h - f_h' \delta r+\frac{f_h'}{2}r_h (C_{\Theta} - 1 ) +\morder{r-r_h}  \,, &\\
 \delta \hat \phi &= - \phi_h' \delta r+\frac{\phi_h'}{2}r_h (C_{\Theta} - 1 ) +\morder{r-r_h} \,,&&&
\end{align}    
\end{subequations}
\rl{where 
$\delta r = -\frac{1}{2}\int_{r_b}^{r_h}d\tilde r \left(\Theta(\tilde r)+\Gamma(\tilde r)\right)-r_b(C_\Theta-1)/2$.} We will also need the shift of the horizon location, which can easily be worked out from the variation of the blackening factor as
\be
 \delta r_h = - \frac{\delta \hat f}{f_h'}\bigg|_{r=r_h} = \delta r - \frac{r_h}{2}  (C_{\Theta} - 1 ) -\frac{1}{2A_h'}h_h \,.
\ee
The full variation of the horizon values of $A$ and $\phi$ are found to be
\be \label{eq:deltaAhfinal}
\delta A_h = A_h' \delta r_h + \delta\hat A \big|_{r=r_h} = - \frac{1}{2} \,,\qquad \delta \phi_h = \phi_h' \delta r_h + \delta\hat \phi \big|_{r=r_h} = -\frac{\phi_h'h_h}{2A_h'} \, . 
\ee
Note the cancellations between the first and second terms of $\delta A_h$ and $\delta \phi_h$. Now, using these expressions we can express the quantity that appears on the RHS of (\ref{eq:GPRbulkviscosity}) neatly as
\be \label{finalvar}
 \frac{\phi_h'h_h}{A_h'}   =   \frac{\delta \phi_h}{\delta A_h} \, .
\ee
This means that we can express the bulk viscosity entirely in terms of horizon data as, 
\begin{align} \label{eq:GPRbulkviscosity2}
    \zeta  = \frac{\mathcal{N}}{9}\,e^{3 A_h} \left(\frac{\delta \phi_h}{\delta A_h}\right)^2  ~~ .
\end{align}
We can further simplify this expression by considering what $\delta A_h$ really means. The charge $\rho$ is only affected by the transformations~\eqref{symmvars1} so that the total variation is $\delta \rho/\rho = -3/2$.
Because $s \sim \exp(3 A_h)$, comparing with~\eqref{eq:deltaAhfinal} we see that $s$ and $\rho$ transform in the same way and therefore the ratio $q  = \rho / s $ is invariant. 
Consequently, one has
\begin{align} \label{eq:densityentropyvariation}
   \frac{1}{3}  \frac{\delta \phi_h}{\delta A_h}   =  s \frac{\partial \phi_h  (s ,q )}{\partial s }  =  s  \frac{\partial \phi_h  (s , \rho  )}{\partial s }  + \rho   \frac{\partial \phi_h  (s , \rho  )}{\partial \rho  }   ~~ .
\end{align}
Finally, from Eq.~\eqref{eq:GPRbulkviscosity} we obtain an expression for the bulk viscosity given by 
\begin{align} \label{eq:EObulkviscosity}
    \zeta   = \frac{  s     }{4 \pi   } \left(s  \frac{\partial  \phi_h }{ \partial  s } 
  + \rho  \frac{\partial  \phi_h }{ \partial  \rho }\right)^{2} ~~ .  
\end{align}
This result agrees with the expression that was found by Eling and Oz \cite{Eling:2011ms} and which was numerically shown to be equivalent to Eq.~\eqref{eq:GPRbulkviscosity} in Ref.~\cite{Buchel:2011wx}. Derivation of \cite{Eling:2011ms} followed from the Raychaudhuri equation near the horizon. Our derivation instead follows from a reinterpretation of the holographic holographic Kubo formula for bulk viscosity in terms of variations of the background fields.  

\section{Anisotropic transport induced by magnetic field}
\label{sec:magneticviscosity}

\subsection{Anisotropic viscosities with the GPR method}\label{aniostvisc}

After the warm up in the previous section, we now focus on the main interest of this paper: holographic transport in anisotropic states. In particular we assume that the anisotropy is brought by the presence of an external magnetic field, which we holographically model by the spatial components of a bulk gauge field. In the presence of a background magnetic field $B^{\mu}$ and when the field theory enjoys parity and time-reversal symmetry, the leading order dissipative correction to stress tensor reads \cite{Hernandez:2017mch,Armas_2019}
\begin{align}  \label{eq:stressmagnet}
\begin{split}
    T_{(1)}^{\mu \nu}  = &  - 2 \eta_{\perp} \Pi^{\mu \nu \alpha \beta } \partial_{\alpha} u_{\beta }   - 2  \eta_{\parallel} b^{(\mu } \Pi^{ \nu) \alpha}  b^{ \beta} \partial_{\alpha} u_{\beta } -  \zeta_{\perp } \Pi^{\mu \nu } \Pi^{ \alpha \beta } \partial_{\alpha} u_{\beta }  \\   & -  \zeta_{\times } (\Pi^{\mu \nu } b^{ \alpha} b^{ \beta } + b^{ \mu} b^{ \nu } \Pi^{\alpha \beta  }  )\partial_{\alpha} u_{\beta }   -  \zeta_{\parallel }  b^{ \alpha} b^{ \beta }  b^{ \mu} b^{ \nu }   \partial_{\alpha} u_{\beta }  ~~  .  
\end{split}
  \end{align}
  where we defined the projectors used the definitions
\begin{align}
 \Pi^{\mu \nu \alpha \beta  } = 
  \Pi^{\mu  ( \alpha }  \Pi^{\beta ) \nu   }-  \frac{1}{2}  \Pi^{\mu \nu }  \Pi^{\alpha \beta  }     ~~ , ~~  \Pi^{\mu \nu } =  \Delta^{\mu \nu }  - b^{\mu} b^{\nu}  ~~ , ~~  b^{\mu} = B^{\mu} / |B|      ~~   , 
\end{align}
with $\Delta_{\mu\nu}$ being the projector transverse to fluid velocity. For the charge current we have 
\begin{align}
    J_{(1)}^{\mu } =   - \sigma_{\parallel} b^{\mu } b^{\nu} T \partial_{\nu} \frac{\mu}{T}  - \sigma_{\perp} \Pi^{\mu \nu } T \partial_{\nu} \frac{\mu}{T} ~~ . 
\end{align}The following constraints on these transport coefficients arise from positivity of entropy production \cite{Hernandez:2017mch}
\begin{align} \label{eq:secondlawwww}
        \eta_{\perp }  \geq 0 ~~ , ~~    \eta_{\parallel }  \geq 0 ~~ , ~~    \zeta_{\perp }  \geq 0 ~~ , ~~   \zeta_{\parallel  }  \geq 0   ~~ , ~~   \zeta_{\perp }    \zeta_{\parallel  }  \geq \zeta_{\times}^2 ~~,~~ \sigma_{\parallel} \geq 0~~,~~ \sigma_{\perp} \geq 0  \,. 
\end{align}
  Holographic values (at infite `t Hooft coupling and infinite $N$) of magnetic shear viscosities  $\eta_{\perp }$ and $\eta_{\parallel}$ were obtained in Ref.~\cite{Jain:2015txa}
  \begin{align} \label{eq:magneticshearviscosities}
\frac{ \eta_{\perp }}{s} = \frac{1}{4 \pi } ~~ , ~~ \frac{\eta_{\parallel}  }{s}  =    \frac{1}{4 \pi } \frac{g_{33}}{g_{11}}  \Big|_{r= r_h } ~~,  
  \end{align}
  where the $z$-direction points along the magnetic field. 
  
  We will now compute the bulk viscosities $\zeta_{\parallel}$, $\zeta_{\perp}$ and $\zeta_{\times}$.  We will do this first by deriving the associated holographic Kubo formulas, following the GPR method, by modifying the computation in Sec.~\ref{sec:gubsergublk} to include  a background magnetic field. We will then generalize the background variation method in Sec.~\ref{sec:GPREOconnection} to this case.

  A background magnetic field can be incorporated in the holographic action as follows,
\be \label{eq:action}
 S = \mathcal{N} \int \sqrt{-g}\left[R -\frac{1}{2}
 (\pa \phi)^2 - V(\phi) + \mathcal{L}_\mathrm{matter}(\phi,F)\right]  ~~ , 
\ee
where for the matter sector we simply take the Maxwell term given by
\be
  \mathcal{L}_\mathrm{matter}(\phi,F) = -\frac{1}{4}Z(\phi) F^2 ~~ , 
\ee
We consider a background magnetic field and chemical potential.
This necessitates an anisotropic metric 
\begin{equation} \label{eq:metricansatz0}
    ds^2=e^{2A(r)}\left(-f(r)dt^2+e^{2W(r)}(dx^2+dy^2)+dz^2+\frac{dr^2}{f(r)} \right) ~~ . 
\end{equation}
The gauge field is given by the ansatz  
\begin{equation}
    A_\mu=\{\Phi(r),-\frac{yB}{2},\frac{xB}{2},0,0\} ~~ , 
\end{equation}
where the chemical potential is given by the source of the temporal component $\Phi$, i.e., $\Phi(r)|_\mathrm{bdry} = \mu$.
When the above ans\"atze are substituted into Einstein equations
\begin{subequations}
\begin{align}
  & R_{\mu\nu}-\frac{1}{2}g_{\mu\nu}R  = T_{\mu \nu } ~~ , \\ 
T_{\mu \nu }     & = \left(\frac{1}{2}\partial_\mu\phi\partial_\nu\phi-\frac{1}{4}g_{\mu\nu}(\partial \phi)^2+\frac{1}{2}g_{\mu\nu}V[\phi]\right) + \frac{Z[\phi]}{2}\left(4{F_{\mu}}^{\lambda}F_{\nu\lambda}-g_{\mu\nu}F_{\lambda\kappa}F^{\lambda\kappa}\right)  \label{eq:stresstensor}
\end{align}      
\end{subequations}
one obtains the following system of differential equations:
\begin{equation}
A''-A'^2+\frac{2}{3}(W''+W'^2)=-\frac{1}{6}\phi'^2,\label{ee1}
\end{equation}
\begin{equation}  
\left(W'e^{3A+2W}f\right)'=-\frac{2B^2Z[\phi]}{e^{2W-A}},\label{ee2}
\end{equation}
\begin{equation}
   \frac{f''}{f}+2A''+\frac{f'}{f}\left(5A'+2W'\right)+2A'\left(3A'+2W'\right)-\frac{2e^{2A}V[\phi]}{3f}-\frac{8\Phi'^2Z[\phi]}{3fe^{2A}} =\frac{4B^2Z[\phi]}{3fe^{2(A+2W)}} ,\label{ee3}
\end{equation}
\begin{equation}
     3A'\left(4A'+\frac{f'}{f}+4W'\right)+2W'\left(\frac{f'}{f}+W'\right)-\frac{1}{2}\phi'^2-\frac{e^{2A}V[\phi]}{f}+\frac{2\Phi'^2Z[\phi]}{fe^{2A}} =-\frac{2B^2Z[\phi]}{fe^{2(A+2W)}}.\label{ee4}
\end{equation}
Let us analyze these equations in some detail. For the moment we do not want to specify the UV boundary conditions. We will only assume that there is a UV boundary at $r_b$ where the conformal factor $A(r)$ becomes large. In fact, choose the scale function $A$ to monotonically decrease from boundary to interior and everywhere satisfy $A'<0$ which is consistent with the requirement that $A$ itself can be used as a holographic coordinate.

On the other hand Einstein's equation (\ref{ee2}) leads to the inequality $(W'e^{(3A+2W)}f)'<0$. This follows from the fact that $Z[\phi]$ in the action should be positive definite which itself follows from the requirement of a positive definite kinetic energy of the bulk gauge field. In addition, for asymptotically AdS boundary one needs that $W\to 0$ as $r\to r_b$. Finally, the blackening factor $f>0$ everywhere outside the horizon, that is $r_b\leq r<r_h$ and it goes to 1 on the boundary. The aforementioned inequality guarantees that $W'\exp(3A+2W)f \to 0^+$ as $r\to r_h$ which, in turn, requires that $W$ is a monotonically increasing function everywhere outside the horizon: 
\be\label{monincW}
W'(r) >0\,, \qquad r_b\leq r < r_h\, .
\ee
This equation will be very useful in proving inequalities below that will restrict transport properties. 

In fact the metric functions satisfy more restrictions. It is straightforward to obtain a modified version of the {\em holographic a-theorem} in the presence of anisotropy from Einstein's equations, see e.g. \cite{Giataganas:2017koz}, 
\be\label{anath}
\frac{d}{dr} \left\{(A'+\frac23 W')e^{-A+\frac23 W}\right\}\leq 0\, . 
\ee
Furthermore, assuming asymptotically AdS boundary, one can show that $A'+\frac23 W'\to -\infty$ at the boundary. Let us assume $r_b=0$ as usual. Asymptotically AdS means that $A'\to -1/r$. Solving (\ref{ee2}) near the boundary, with the additional assumption that $Z$ approaches a constant, one finds $W \to W_0 r^4 \log r/\ell$ with $W_0<0$. That is $A'+\frac23 W'<0$ everywhere. Multiplying this with $W'$, which we showed to be positive definite above, we find that $A'<0$ which is consistent with the discussion above.  

The we consider the fluctuated metric 
\begin{equation} \label{eq:metricansatz}
    ds^2=e^{2A(r)}\left(-f(r)dt^2+e^{2W(r)}(dx^2+dy^2)+dz^2+\frac{dr^2}{f(r)} + 
 H_{MN}(r , t ) dx^M dx^N \right) ~~ , 
\end{equation}
where the scalar sector metric fluctuations are given by
\begin{align}
\label{eq:wfdefs}
\begin{aligned}
    H_{11} &= H_{22} =  e^{2 W (r)} h_{\perp }(r)  e^{-i\omega t} \ ,
    \\
     H_{33} &  =  h_{\parallel} (r) e^{-i\omega t}  \ , \\ 
      H_{tt} &= - f(r) (\Theta(r) + h_{\parallel}(r)) e^{-i\omega t}  \ .
    \end{aligned}
&& 
\begin{aligned}
     H_{rt} &= h_{rt}(r) e^{-i\omega t}  \\
 H_{rr} &= f(r)^{-1}(\Gamma(r) + h_{\parallel}(r))e^{-i\omega t}   \\
  & 
\end{aligned}
\end{align}
We see that we now have two spatial scalar fluctuations $ h_{\perp }(r)$ and $ h_{\parallel }(r)$, which 
satisfy coupled 
fluctuation equations, unlike was the case for the isotropic $h(r)$ whose decoupled equation was given in Eq.~\eqref{hfluct}. However, to obtain an analytic result for the bulk viscosity, only the near horizon behavior will be relevant. For this purpose, one can construct a linear combination of fluctuations for which a decoupling indeed occurs near the horizon. Specifically, the following fluctuations which we call $\Delta(r)$ and $\Sigma(r)$ 
\begin{equation}\label{DS1inv}
    \Delta(r)=\frac{1}{3}\left[h_{\parallel }(r)-\frac{A'}{A'+W'}h_{\perp}(r)\right] \qquad \text{and} \qquad \Sigma(r)=\frac{h_{\parallel}(r)+2h_{\perp}(r)}{3} ~~  ,
\end{equation}
satisfy the equations of motion
\begin{subequations}\label{fluceqsd}
\begin{align}
    \Delta''  & + \left( \frac{f^{\prime}}{f}   + \mathcal{K}_{\Delta}^{(1)} \right)  \Delta'+  \left( \frac{\omega^2}{f^2 } + \mathcal{K}_{\Delta}^{(2)} \right) \Delta+ \mathcal{K}^{(3)}_{\Delta} \Sigma  =0  ~~ , \\ 
    \Sigma''   & +  \left( \frac{f^{\prime}}{f}   +\mathcal{K}^{(1)}_{\Sigma} \right)  \Sigma'+ \left( \frac{\omega^{2}}{f^2 } + \mathcal{K}^{(2)}_{\Sigma}  \right)  \Sigma+ \mathcal{K}^{(3)}_{\Sigma}\Delta  =0 ~~  
, 
\end{align}    
\end{subequations}
where $\mathcal{K}_{\Delta}^{(1)}, \mathcal{K}_{\Delta}^{(2)}, \mathcal{K}_{\Delta}^{(3)}, \mathcal{K}_{\Sigma}^{(1)}, \mathcal{K}_{\Sigma}^{(2)}, \mathcal{K}_{\Sigma}^{(3)}$ are coefficients that depend on the background fields and are subleading near the horizon. We now have two spatial fluctuations in the scalar sector, and we must choose which one to turn on at the boundary. To distinguish these choices, we define
\begin{align}
  \lim_{r \rightarrow r_b}  \begin{pmatrix}
       \Delta (r )  \\ 
 \Sigma (r )   \end{pmatrix} = \mathbf{v}^I  ~~ ,  ~~  , 
\end{align}
where $I$ is a label. We consider two cases, namely
\begin{align} \label{eq:boundaryconditions}
    \mathbf{v}^{\Delta} =   \begin{pmatrix}
       1  \\ 
0   \end{pmatrix} ~~ , ~~   \mathbf{v}^{\Sigma} =    \begin{pmatrix}
       0  \\ 
1  \end{pmatrix}  ~~ . 
\end{align}
Near the horizon, we can solve Eq.~\eqref{fluceqsd} analytically and impose infalling boundary conditions, leading to
\begin{subequations}\label{eq:horizonsolution}
\begin{align} 
\Sigma^{I } (r )  & \approx \Sigma_h^{I }  (r_h - r)^{-  \frac{i \omega}{4 \pi T }}  \approx \Sigma_h^{I }  \left[ 1 -  \frac{i \omega}{4 \pi T }  \log (r_h- r)  \right] ~~ ,  \\ 
\Delta^{I } (r )  & \approx \Delta_h^{I }  (r_h - r)^{-  \frac{i \omega}{4 \pi T }}  \approx \Delta_h^{I }  \left[ 1 -  \frac{i \omega}{4 \pi T }  \log (r_h- r)  \right]  
 ~~ ,  
\end{align}    
\end{subequations}
where $\Sigma_h^{I }$ and $ \Delta_h^{I }$ are coefficients that can be obtained by numerically solving the fluctuation equations from the boundary to the horizon for $\omega =0 $. Again performing integration by parts, we can write our quadratically expanded Lagrangian $  \mathcal{L}_{IJ}^{(2)}$ 
evaluated on the fluctuations satisfying the 
boundary conditions $I,J$ as  \cite{Kaminski:2009dh}
\begin{align} \label{eq:quadraticaction}
    \mathcal{L}_{IJ}^{(2)} &= \partial_r J_{IJ} +  \vec{h}^{* T} \left( \frac{\partial \mathcal{L}_{IJ}}{ \partial \vec{h}^* } - \partial_r \frac{\partial \mathcal{L}_{IJ}}{ \partial  \vec{h}^{* \prime }   }  \right) ~~  , ~~  \vec{h} = \{ \Delta ,  \Sigma   , \Theta  , \Gamma     \}  ~~ . 
\end{align}
Here the index $I$ ($J$) refers to the boundary condition for the conjugated (unconjugated) fields, and
\begin{equation}
    J_{IJ}= \frac{27e^{3A+2W}f(A'+W')^2}{(3A'+2W')^2}\Delta_I^*(r) \Delta_J'(r)+\frac{9}{4}\frac{e^{3A+2W}f\phi'^2}{(3A'+2W')^2}\Sigma_I^*(r) \Sigma_J'(r) ~~ .
\end{equation}
From Eq.~\eqref{eq:quadraticaction} we can obtain the retarded Green's function
\begin{align}
\text{Im }  G^R_{IJ} (\omega )  = -  \mathcal{N} \mathcal{F}_{IJ}~~  ,  ~~ \mathcal{F}_{IJ} = - \text{Im } J_{IJ}  ~~ . 
\end{align}
Taking $\mathcal{F}_{IJ}$ to the horizon, we obtain
\begin{align} 
\begin{split}    
 \mathcal{N} \mathcal{F}_{IJ} &= 
  \frac{\omega s}{\pi}\frac{1}{(3A'_h+2W'_h)^2}\left[\frac{9}{16}\phi_h^{\prime 2}\mathrm{Re} \left( \,(\Sigma^I_h )^*\Sigma_h^J \right) +\frac{27}{4}(A'_h+W'_h)^2\mathrm{Re} \left( \,(\Delta_h^I)^*\Delta_h^J \right)\right]\bigg|_{r\rightarrow r_h}\!\! + \morder{\omega^2} .\label{F}
\end{split}
\end{align}
The on-shell action only depends on the boundary values of  $\Sigma$ and $\Delta$, and taking derivatives with respect to them 
yields the following dictionary in terms of the components of the boundary stress tensor,
\begin{align}
  \frac{1}{V_4}\frac{\partial S_\mathrm{on-shell}}{\partial \Sigma(r_b)}  
    = \frac{1}{2} \sum_{k=1}^3T_{kk}  ~~ , ~~    
   \frac{1}{V_4}\frac{\partial S_\mathrm{on-shell}}{\partial \Delta(r_b)}
   = T^{33}-(T^{11}+T^{22})/2 ~~  , 
\end{align}
where $V_4$ is the volume of spacetime. By using the expansion of the boundary stress tensor in Eq.~\eqref{eq:stressmagnet}, we can find the Kubo formulas for the bulk viscosities
\begin{equation}
 \zeta_\perp = -\lim_{\omega \to 0}\frac{1}{\omega} \,\mathrm{Im}\, G_{\mathcal{O}_1\mathcal{O}_1}(\omega) \ , \qquad
\zeta_\times = -\lim_{\omega \to 0}\frac{1}{\omega} \,\mathrm{Im}\, G_{\mathcal{O}_1T^{33}}(\omega) \ , \qquad 
\zeta_\parallel = -\lim_{\omega \to 0}\frac{1}{\omega} \,\mathrm{Im}\, G_{T^{33}T^{33}}(\omega) \ ,
\end{equation}
where $\mathcal{O}_1 = (T^{11}+T^{22})/2$.
Applying these formulas to our case, we identify the coefficients of the corresponding terms i.e. the anisotropic bulk viscosities with the following specific combinations of the flux of bulk fluctuations 
\begin{subequations}  \label{eq:zetformulae}
\begin{align}
    \zeta_{\perp } & =  \mathcal{N} \lim_{\omega \to 0}\frac{1}{\omega} \,\left(  \frac{\mathcal{F}_{\Delta\Delta}}{9}-\frac{4 \mathcal{F}_{\Sigma \Delta}}{9}+\frac{4 \mathcal{F}_{\Sigma \Sigma}}{9} \right)  ~~ ,   \\ 
     \zeta_{\times} & =  \mathcal{N}\lim_{\omega \to 0}\frac{1}{\omega} \, \left( -\frac{2}{9} \mathcal{F}_{\Delta\Delta}  + \frac{2}{9} \mathcal{F}_{\Sigma \Delta}+\frac{4}{9} \mathcal{F}_{\Sigma \Sigma}\right) ~~ , \\ 
     \zeta_{\parallel}    & =  \mathcal{N}\lim_{\omega \to 0}\frac{1}{\omega} \,\left(\frac{4}{9} \mathcal{F}_{\Delta\Delta}+ \frac{8}{9} \mathcal{F}_{\Sigma \Delta}+ \frac{4}{9} \mathcal{F}_{\Sigma \Sigma}  \right)  ~~ .
\end{align}  
\end{subequations}
These are the holographic Kubo formulae which will serve as the starting point for our derivation in terms of background variations below. 
\subsection{Transforming fluctuations to background variations: finite magnetic field}
The GPR result in Eq.~\eqref{eq:zetformulae} depend on the horizon values of fluctuations. In this section, we will see that, in analogy to Sec.~\ref{sec:GPREOconnection}, we can express these results solely in terms of horizon values of background fields by rewriting fluctuations with vanishing frequency in terms of variations of the background. As before, we will relate fluctuations to background variations in the gauge where dilaton fluctuation is turned off, yielding the following relations
\begin{subequations}  \label{eq:mapping3}
\begin{align}     \label{mapping1987987}
 \chi(r) &= 0 \,,&\\
 h_{\parallel}(r) &= 2 \delta A(r) - 2 \frac{A'(r)}{\phi'(r)}\, \delta \phi(r) \,,&\\
  h_{\perp}(r) &= 2 \delta A(r) +2 \delta W(r) - 2 \frac{A'(r)}{\phi'(r)}\, \delta \phi(r)- 2 \frac{W'(r)}{\phi'(r)}\, \delta \phi(r) \,,& \\ 
 \Theta(r) &=
 \frac{\delta f(r)}{f(r)} 
 - \frac{f'(r)}{f(r)\phi'(r)}\, \delta \phi(r) \,,&\\
  \Gamma(r) &= 
  -\frac{\delta f(r)}{f(r)} 
  + \frac{f'(r)}{f(r)\phi'(r)}\, \delta \phi(r)  - 2 \frac{\delta\phi'(r)}{\phi'(r)}+ 2 \frac{\phi''(r)\delta\phi(r)}{\phi'(r)^2}\,, &
\label{mapping2987987}
\end{align}
\end{subequations}
\rl{for the wave functions defined in~\eqref{eq:wfdefs}.}
On the other hand, we will also need global symmetries of the background --- which can be obtained from symmetries of the action --- in order to switch off the boundary values of the sources as above. At the level of background solutions, these symmetries are \cite{Gursoy:2020kjd}
\begin{subequations}
    \begin{align} \label{eq:symmBfirst}
 r &\to r +r_0 \,, &&&\\
 \label{eq:symmBsecond}
 A &\to A + \log \Lambda \,,&\qquad r &\to \Lambda r \,,&\qquad B &\to \Lambda^2 B\,,&\qquad \rho &\to \Lambda^3 \rho\,,&\\ 
 f &\to f_0 f\,,&\qquad A &\to A + \frac{1}{2}\log f_0\,,&\qquad B &\to f_0 B\,,&\qquad \rho &\to f_0^{3/2} \rho\,,&\\
 B &\to \Lambda_\perp^2 B \,,&\qquad W &\to W +\log\Lambda_\perp\,,&\qquad   \rho  &  \to \Lambda_\perp^2 \rho \, ,   &
  \label{eq:symmBlast}
\end{align}
\end{subequations}
where each line corresponds to a separate class of symmetry, hence there are four separate classes of symmetry transformations. The infinitesimal version of these symmetry parameters are
\be
r_0 = \eps_r\,,\qquad  \Lambda = 1+\eps_\Lambda \,,\qquad f_0 = 1+\eps_f\,,\qquad \Lambda_\perp = 1+\eps_\perp\,, 
\ee
and they generate the following variation of the background:
\begin{align}\label{symmvars112}
\begin{aligned}
    \delta_\epsilon A &= \epsilon_\Lambda+ (r \epsilon_\Lambda +\epsilon_r) A' +\frac{\epsilon_f}{2}\,,  \\ 
\delta_\epsilon \phi &=  (r \epsilon_\Lambda +\epsilon_r) \phi'\,, \\ 
\delta_\epsilon W &= \epsilon_\perp + (r \epsilon_\Lambda +\epsilon_r) W' \,,
\end{aligned}
&& 
\begin{aligned}
     \delta_\epsilon f &=  (r \epsilon_\Lambda +\epsilon_r)f' + \epsilon_f f \,, \\ 
      \delta_\epsilon \rho &= (3 \eps_\Lambda + 3 \eps_f/2+ 2\epsilon_\perp) \rho \, ,  \\ 
      \delta_\epsilon B &= \left(\epsilon_f+2\epsilon_\Lambda+2 \epsilon_\perp\right)B \,,
\end{aligned}
\end{align}
We will then perform a symmetry transformation to remove the fluctuations for two boundary conditions labelled by $I$ in Eq.~\eqref{eq:boundaryconditions}. Using Eq.~\eqref{DS1inv}, we find that the fluctuations $\Delta (r)$ and $\Sigma (r)$ at the boundary are given by
\begin{equation}\label{DS1323}
   \lim_{r \rightarrow r_b }   \Delta^I(r)=  \lim_{r \rightarrow r_b }  \left( \frac{1}{3}\left[   h^I_{\parallel}(r)-    h^I_{\perp}(r)\right] \right)  ~~ , ~~  \lim_{r \rightarrow r_b }  \Sigma^I (r)=  \lim_{r \rightarrow r_b }  \left( \frac{  h_{\parallel}^I (r)+2   h_{\perp}^I(r)}{3} \right) ,
\end{equation}
Using Eq.~\eqref{eq:mapping3} and Eq.~\eqref{DS1323}, this relates the variation of the fluctuation to the shifts in the background as
\begin{equation}\label{DS21331}
   \lim_{r \rightarrow r_b } \delta_\epsilon   \Delta^I (r)=   -\frac{2}{3} \epsilon^I_{\perp }   \qquad \text{and} \qquad  \lim_{r \rightarrow r_b }  \delta_\epsilon  \Sigma^I(r)=   2 \epsilon^I_\Lambda+  \epsilon^I_f  + \frac{4}{3} \epsilon^I_{\perp }  ,
\end{equation}
Turning off the fluctuations at the UV boundary then requires
\begin{align} \label{eq:alleps}
\begin{aligned}
    \eps_f^{\Delta } &  = - C_{\Theta}  \,, \\ 
     \eps_f^{\Sigma } &  = - C_{\Theta}  \,,
\end{aligned} && 
\begin{aligned}
    \eps^{\Delta }_\Lambda  & = -  1 + \frac{1}{2} C_{\Theta}   \,, \\ 
    \eps^{\Sigma }_\Lambda  &  = -  \frac{1}{2}+   \frac{1}{2}   C_{\Theta}    \,,
\end{aligned}
&&
\begin{aligned}
    \eps^{\Delta }_\perp   &  =  \frac{3}{2}   ~~ ,  \\ 
     \eps^{\Sigma }_\perp  & =  0    ~~ . 
\end{aligned}
\end{align}
\rl{ As in the previous section, the value of $\eps_r$ is also determined by requiring  the absence of the source term for the generated $\delta \phi$, but its value is not needed in the analysis below.}

In analogy to the warm-up exercise in the previous section next we work out fluctuations near the horizon in order to relate them to horizon values of background variations. In general one has
\begin{align}
\begin{aligned}
     A(r) &= A_h + A_h'(r-r_h) + \morder{(r-r_h)^2}\,, \\ 
      W(r) &= W_h + W_h'(r-r_h) + \morder{(r-r_h)^2}\,,
\end{aligned}
&&
\begin{aligned}
    f(r)& = f'_h(r-r_h) +\morder{(r-r_h)^2}  \,. \\
     \phi(r) &=\phi_h +\phi_h'(r-r_h)+\morder{(r-r_h)^2}  \,.
\end{aligned}
\end{align}     
Let us also denote horizon expansion of the fluctuation equations as
\begin{subequations}  \label{eq:expansions}
\begin{align}
 \Delta^I (r)  & =  \Delta^I_h +  \Delta^{I \prime}_h(r-r_h) + \morder{(r-r_h)^2} \,,  \\ 
  \Sigma^I (r)  & =   \Sigma^I_h +   \Sigma^{I \prime}_h  (r-r_h) + \morder{(r-r_h)^2} \,,\\
 \Theta^I  (r)  & = \frac{\Theta^I_r}{r - r_h } + \Theta^I_h + \Theta_h^{ I \prime}  (r-r_h) + \morder{(r-r_h)^2} \,.    
\end{align}
 \end{subequations}
 $   \Theta_r $ is the only coefficient that must be obtained from the fluctuation equations. From the $rr$-component of the Einstein fluctuation equations, we extract
 \begin{align} \label{eq:nearhorizontheta209109}
    \Theta^I_r   = \frac{3 \Sigma^I_h }{2 (3 A_h^{\prime} + 2 W_h^{\prime})}    ~~. 
\end{align}
Using Eqs.~\eqref{eq:nearhorizontheta209109},~\eqref{eq:mapping3} and~\eqref{eq:alleps} we obtain the total background variations \rl{ at the horizon}
\begin{subequations}
\begin{align}
 \delta\hat A^I &= \frac{3A^{ \prime}_h \Sigma^I_h 
 +6(A^{\prime}_h+W_h')\Delta^I_h }{2( 3A_h'+2W_h') } -A_h' \delta r^I+ \epsilon^I_\Lambda+ r_h \epsilon^I_\Lambda A'_h +\frac{\epsilon^I_f}{2} \,,\\
 \begin{split}
  \delta\hat W^I &=   \frac{3 W_h '\Sigma^I_h 
 -9(A_h'+W_h')\Delta^I_h }{2( 3A_h'+2W_h') }  -W_h' \delta r^I 
 + \epsilon^I_\perp + r_h \epsilon^I_\Lambda W_h' \,,
 \end{split}
  \\
 \delta \hat f^I &= \frac{3 \Sigma^I_h f_h' }{2 (3 A^{\prime}_h + 2 W^{\prime}_h)} - f_h' \delta r^I+  r_h \epsilon^I_\Lambda f_h' +\morder{r-r_h}  \,, \\
 \delta \hat \phi^I &= - \phi_h' \delta r^I+ r_h \epsilon^I_\Lambda \phi_h' +\morder{r-r_h} \,, 
\end{align}   \label{eq:fehiweuhwi111}
\end{subequations}
\rl{ where, following the steps outlined in the previous section,
\be
 \delta r^I = - r_b \eps^I_\Lambda - \frac{1}{2}\int_{r_b}^{r_h}d\tilde r\left(\Theta^I(\tilde r)+\Gamma^I(\tilde r)\right)\,.
\ee
}
The horizon values of $W$, $A$ and $\phi$ change as
\be\label{totvar}
\delta A_h = A_h' \delta r_h + \delta\hat A \big|_{r=r_h}  \,,\qquad \delta W_h = W_h' \delta r_h + \delta\hat W \big|_{r=r_h}  \,,\qquad \delta \phi_h = \phi_h' \delta r_h + \delta\hat \phi \big|_{r=r_h}, 
\ee
where the shift of the horizon location can be computed from variation of the blackening factor as before:
\be\label{varrh}
 \delta r^I_h = - \frac{\delta \hat f^I}{f_h'}\bigg|_{r=r_h} = \delta r^I -   r_h \epsilon^I_\Lambda  -\frac{3 \Sigma^I_h   }{2 (3 A^{\prime}_h + 2 W^{\prime}_h)}  \, . 
\ee
Now we will relate fluctuations at the horizon to variations of the background. As in Sec.~\ref{sec:GPREOconnection}, the variations we consider here keep the ratio $q = \rho / s $ invariant. This means that variation of a general function $g_h (s , q , B ) $ obeys
\begin{align} \label{eq:generalidentity}
    \delta g_h (s , q , B )   = \delta s \frac{\partial g_h  (s , q , B )}{\partial s }  + \delta B \frac{\partial g_h  (s , q , B )}{\partial B } ~~  \,.
\end{align}
All in all, combining Eqs.~(\ref{eq:fehiweuhwi111}), (\ref{totvar}), (\ref{varrh}) and (\ref{eq:generalidentity}) we arrive at the following relations between horizon values of fluctuations and variations of the background
 \begin{subequations} \label{eq:theplugequations}
          \begin{align}
          \frac{ \Sigma^{\Delta}_h \phi_h' }{ (3 A^{\prime}_h + 2 W^{\prime}_h)} &  =  - \frac{2}{3} B \frac{\partial \phi_h  
 (s , q , B )  }{\partial B } ~~ , \\ 
                   \frac{ \Sigma^{\Sigma}_h \phi_h' }{ (3 A^{\prime}_h + 2 W^{\prime}_h)} &   =    s \frac{\partial  \phi_h (s , q , B )   }{\partial s }   +  \frac{2}{3} B \frac{\partial \phi_h (s , q , B )   }{\partial B } ~~ ,        \\  \label{eq:firsteq1} 
 \frac{    3(A_h'+W_h')\Delta^{\Delta}_h }{2( 3A_h'+2W_h') }  
  &  = -   B \frac{  \partial (   A_h (s , q , B ) +   W_h (s , q , B ) ) }{\partial B }+ \frac{1}{2}  ~~  ,  \\    \label{eq:secondeq2}
 \frac{   3 (A_h'+W_h')\Delta^{\Sigma}_h }{2( 3A_h'+2W_h') }  
 &  =   \frac{3}{2}   s \frac{ \partial (   A_h (s , q , B )+   W_h (s , q , B ) )  }{\partial s }   +  B \frac{ \partial (   A_h  (s , q , B )+   W_h  (s , q , B )  )  }{\partial  B }  - \frac{1}{2}  ~~ . 
\end{align}    
\end{subequations}
We can further simplify these equations upon using the following relations 
\begin{subequations}  \label{eq:simplifications}   
\begin{align}
 B\frac{\partial (A_h (s , q , B )  + W_h (s , q , B ))}{\partial B } &=  \frac{B}{3} \frac{\partial W_h (s , q , B )}{\partial B }  ~~ ,   \\ 
 s\frac{\partial (A_h (s , q , B ) + W_h (s , q , B ))}{\partial s }    &= \frac{1}{3} + \frac{1}{3}    s \frac{\partial W_h (s , q , B )}{\partial s }  ~~ ,  \end{align}
\end{subequations}
which follow from the relation 
\begin{align} \label{eq:entropydensity}
s =  4  \pi   \mathcal{N}   \exp(3 A_h + 2 W_h ) ~~ \,,    \end{align}
that constrains the derivative of the combination $3A_h + 2W_h$. 
Combining  Eq.~\eqref{eq:simplifications} with  Eq.~\eqref{eq:densityentropyvariation}, one finds
\begin{subequations} \label{eq:reducedequations}
    \begin{align}
        \frac{ \Sigma^{\Delta}_h \phi_h' }{ (3 A^{\prime}_h + 2 W^{\prime}_h)} &  =  - \frac{2}{3} B \frac{\partial \phi_h  
 (s , \rho  , B )  }{\partial B } ~~ , \\ 
                   \frac{ \Sigma^{\Sigma}_h \phi_h' }{ (3 A^{\prime}_h + 2 W^{\prime}_h)} &   =    s \frac{\partial  \phi_h (s , \rho  , B )   }{\partial s }   + \rho  \frac{\partial  \phi_h (s , \rho  , B )   }{\partial \rho }   +  \frac{2}{3} B \frac{\partial \phi_h (s , \rho  , B )   }{\partial B } ~~ ,    \\ 
        \frac{    3(A_h'+W_h')\Delta^{\Delta}_h }{2( 3A_h'+2W_h') }  
  &  = - \frac{B}{3} \frac{\partial W_h (s , \rho  , B) }{\partial B }+ \frac{1}{2}  ~~  ,  \\ 
 \frac{   3 (A_h'+W_h')\Delta^{\Sigma}_h }{2( 3A_h'+2W_h') }  
 &  =    \frac{s}{2} \frac{\partial W_h (s , \rho  , B) }{\partial s }  +  \frac{\rho }{2} \frac{\partial W_h (s , \rho  , B) }{\partial \rho  }  +  \frac{B}{3} \frac{\partial W_h (s , \rho  , B) }{\partial B }  ~~   . 
    \end{align}
\end{subequations}
Substitution of Eq.~\eqref{eq:reducedequations} into Eq.~\eqref{eq:zetformulae} through Eq.~\eqref{F}, yields
\begin{subequations} \label{eq:viscosoties989821}
\begin{align} 
\begin{split}
    \zeta_{\perp } & = \frac{ s  }{4 \pi }   
  \left(  s\frac{\partial  \phi_h }{\partial s }  +\rho \frac{\partial  \phi_h }{\partial \rho  }    +    B \frac{ \partial  \phi_h}{\partial  B }    \right)^2
+  \frac{ s  }{3 \pi } 
   \left(  s\frac{\partial  W_h }{\partial s }  +\rho \frac{\partial  W_h }{\partial \rho  }  + B \frac{\partial W_h}{\partial B} - \frac{1}{2}   \right)^2     ~~ ,  
\end{split}
     \\ 
 \begin{split}    
     \zeta_{\times} & = \frac{ s  }{4 \pi }  \left(s   \frac{\partial \phi_h}{ \partial s}  +\rho \frac{\partial  \phi_h }{\partial \rho  }    +    B \frac{ \partial  \phi_h}{\partial  B }    \right) \left( s\frac{\partial  \phi_h }{\partial s }  +\rho \frac{\partial  \phi_h }{\partial \rho  }     \right)  \\  & 
     +  \frac{  s }{ 3 \pi } 
 \left( B \frac{\partial W_h  }{\partial B}  + s\frac{\partial  W_h }{\partial s }  +\rho \frac{\partial  W_h }{\partial \rho  }- \frac{1}{2}  \right)     \left(  s\frac{\partial  W_h }{\partial s }  +\rho \frac{\partial  W_h }{\partial \rho  }   +1  \right)  ~~ , 
 \end{split} \\ 
     \zeta_{\parallel}    & =   \frac{  s  }{4 \pi }   \left(  s \frac{ \partial  \phi_h}{\partial s }  + \rho  \frac{  \partial  \phi_h}{\partial \rho }  \right)^2      + \frac{ s  }{3 \pi }\left( s\frac{\partial  W_h }{\partial s }  +\rho \frac{\partial  W_h }{\partial \rho  }  + 1  \right)^2  ~~ .
\end{align}    
\end{subequations}
These relations constitute our final result for the anisotropic bulk viscosities induced by an external magnetic field. The results are (semi-)analytic, in the sense that values of transport coefficients directly follow if one knows the background analytically. We also note that, as in the cases of isotropic bulk viscosity and anisotropic shear viscosities, the transport coefficients are expressed purely in terms of horizon data. When taking the non-magnetic limit for Eq.~\eqref{eq:viscosoties989821}, it is important to note that the anisotropic shear viscosities introduced in Eq.~\eqref{eq:stressmagnet} are only traceless for the transverse tensor $\Pi^{\mu \nu }$. This means that part of the isotropic shear viscosity, which is traceless for the tensor $\Delta^{\mu \nu }$, will be contained by the dilaton-independent part of the anisotropic bulk viscosities in Eq.~\eqref{eq:viscosoties989821}. Specifically, taking the non-magnetic limit yields
\begin{subequations} \label{eq:viscosoties9898223321}
\begin{align} 
\begin{split}
  \lim_{B \rightarrow 0 }  \zeta_{\perp } & = \zeta   +  \frac{1}{3} \eta   ~~ ,  
\end{split}
     \\ 
 \begin{split}    
   \lim_{B \rightarrow 0 }    \zeta_{\times} & = \zeta -  \frac{2}{3} \eta  ~~ , 
 \end{split} \\ 
   \lim_{B \rightarrow 0 }    \zeta_{\parallel}    & =\zeta +   \frac{4}{3} \eta ~~ .
\end{align}    
\end{subequations}
Finally, it is straightforward to check that all the constraints that arise from the Second Eq.~\eqref{eq:secondlawwww} are automatically satisfied for Eq.~\eqref{eq:viscosoties989821}.
\subsection{Anisotropic conductivities induced by magnetic field and universal relations}
We can easily generalize our analysis to compute electric conductivity in the presence of an external magnetic field. As the magnetic field breaks isotropy, as above, there are separate components parallel and transverse to $B$. Holographically the situation is modelled by considering a gauge field 
\begin{equation}
    A_\mu=\{\Phi(r),-\frac{y B}{2},\frac{x B}{2},0,0\}  +  a_{\mu} ~~ , 
\end{equation}
where $a_{\mu}$ is a small fluctuation. Metric fluctuations decouple completely from fluctuations of the gauge field. Focusing solely on the gauge field fluctuations then, we expand the Lagrangian quadratically in the sources as 
\begin{align}
    \mathcal{L}^{(2)} = e^{A} Z (\phi )   \left( |a_1^{\prime}|^{ 2 } - \frac{\omega^2}{f^2 } |a_1|^2 + e^{2 W}  |a_3^{\prime}|^{ 2 }-\frac{\omega^2}{f^2 } e^{2 W} | a_3|^2  \right) + ... ~~   .  \label{eq:complexvalued}
\end{align}
This can be put in the standard form
\begin{align}
    \mathcal{L}^{(2)} = \partial_r \mathcal{J} +   a_{\mu}^{*} \left( \frac{\partial \mathcal{L}}{ \partial a_{\mu}^* } - \partial_r \frac{\partial \mathcal{L}}{ \partial a_{\mu}^{* \prime }   }  \right) ~~  , ~~   , 
\end{align}
where the $\mathcal{J}$-term is the only part that is non-vanishing on-shell. This term can be related to the retarded Green's function as \cite{Herzog_2003,Son_2002}
\begin{align}
\text{Im }  G^R (\omega )  =  - \mathcal{N}\mathcal{F} ~~  ,  ~~ \mathcal{F} = - \text{Im }  \mathcal{J} ~~ , 
\end{align}
where $G^R$ is the thermal retarded Green's function of the electric current $J_i$ in the boundary field theory 
\begin{align} \label{eq:greensfunctionrelation11}
     G^R (\omega )   =  - i \int dt dx^3 e^{i \omega t }  \langle [  J_{i} (t , \vec{x}) 
 ,  J_{i  } (0,0 )  ] \rangle ~~ . 
\end{align}
Here we find from the 
flux $\mathcal{F}$ from the holographic action \eqref{eq:action} as 
    \begin{align}
 \label{eq:Fterm}
\mathcal{F} = - \frac{i }{2} \lim_{r \rightarrow r_h } 
 Z (\phi ) \exp( A )  f \left[  \left( a_1' a_1^{*} - a_1^{* \prime } a_1 \right)  +  \exp(2 W ) \left( a_3' a_3^{*} - a_3^{* \prime } a_3 \right)  \right]   ~~ . 
    \end{align}
Fluctuation equations for $a_1$ and $a_3$ are of the form
\begin{align} \label{eq:equatoin}
   a_i''  & + \left( \frac{f^{\prime}}{f}   + \mathcal{K}  \right)  a_i'+  \frac{\omega^2}{f^2 }  a_i   =0  ~~ , \end{align}
   with $\mathcal{K}$ is some function depending on the background fields, which we can solve near the horizon as
   \begin{align}
      a_i  =   c \left[1 - \frac{i \omega }{4 \pi T } \log( r_h - r  )  \right]  ~~. 
   \end{align}
For $\omega=0$, it follows from Eq.~\eqref{eq:equatoin} that we have a trivial flow which allows us to obtain $c=1$ and we can thus extract from Eq.~\eqref{eq:Fterm} the conductivities
\begin{align}
 \label{condeq}   \sigma_{\parallel} =  \mathcal{N} \exp(A_h + 2 W_h) Z(\phi_h)  ~~ , ~~ \sigma_{\perp }  =    \mathcal{N} \exp(A_h  ) Z(\phi_h) ~~ . 
\end{align}
Interestingly, comparison to Eq.~\eqref{eq:magneticshearviscosities} leads to an interesting relation between the magnetic shear viscosities and conductivities:
\begin{align}
    \frac{\eta_{\parallel}}{\eta_{\perp}} =  \frac{\sigma_{\perp}}{\sigma_{\parallel}} ~~ .  
\end{align}
This relation between anisotropic viscosities and conductivities were already observed\footnote{This observation was in a different setting where the anisotropy gave rise to a \rl{different kind of boost symmetry breaking, namely along the direction of anisotropy as opposite to on the transverse plane \cite{Jain:2015txa}. This leads to a different relation with one side of this equation being inverted.}} in \cite{Landsteiner_2016,Baggioli_2018}. 

Furthermore we find by substituting the analytical expressions \eqref{eq:magneticshearviscosities} 
 and \eqref{condeq} for the transport coefficients in the above ratio, that one can express it in terms of horizon data: 
\begin{align}
\frac{\eta_{\perp}}{\eta_{\parallel}} =  \frac{\sigma_{\parallel}}{\sigma_{\perp}}=e^{2W_h}~~. 
\end{align}
This simple expression is quite useful and leads to a universal inequality. Using the fact that $W$ is a monotonically increasing function that is positive definite everywhere, see Eq.~(\ref{monincW}) and above, we find the {\em universal inequality} 
\begin{align}
\frac{\eta_{\perp}}{\eta_{\parallel}} =  \frac{\sigma_{\parallel}}{\sigma_{\perp}}\geq1. 
\end{align}
We note that these type of inequalities in ratios of different components of transport coefficients are ubiqitious in holographic QCD theories, see for example \cite{Gursoy:2010aa} for a similar inequality for the ratio of Langevin diffusion constants. 
\section{Transport from EO method}\label{sec:EO}

In this section we will use the EO approach, introduced in \cite{Eling:2010hu,Eling:2011ms}, to compute the entropy production current of a plasma by considering Einstein equations near the horizon. This will allow us to reproduce the results for bulk and shear viscosities in anisotropic plasmas shown in the previous section, however we will also go beyond this and extend the result to a class of higher derivative theories.  We work with the following generic two derivative action with a Gauss-Bonnet term\footnote{Note that we are choosing the normalization to be $\mathcal{N}= \frac{1}{16 \pi G_N} $, with $G_N = 1$. }
\begin{align}
 S&=  \int \frac{d^5x \sqrt{-g}}{16 \pi} \left[ R - \frac{\left(\partial \phi\right)^2}{2} - V(\phi) - \frac{Z(\phi)F^2}{4}  + \frac{\lambda}{2} \left(R^2-4 R^{A B} R_{AB} +R^{ABCD}R_{CDAB} \right) \right] 
\end{align}
where $\lambda$ is the Gauss-Bonnet coupling constant.
 The equations of motion that follow from this action read
\begin{align}
    \nabla_M \left[ Z F^{M N} \right] &=0 \\ 
    E_{AB} \equiv R_{AB}- \frac{1}{2} g_{AB} R + 2\lambda \mathcal{H}_{AB}- \mathcal{T}_{AB} &=0 
\end{align}
where $\mathcal{H}_{AB}$ is the Lanczos tensor given by
\begin{align}
    \mathcal{H}_{AB} &= \frac{1}{2} \left( R R_{AB} - 2 R_{ACBD}R^{DC} - 2 R_{AC} R^C{}_B + R_{A L}{}^{CD} R_{CDB}{}^L \right)   \\
    &- \frac{1}{8} g_{AB}  \left(R^2-4 R^{CD} R_{CD} +R^{EFCD}R_{CDEF} \right)  \nonumber \, , 
\end{align}
and $\mathcal{T}_{AB}$ is the energy momentum tensor given by
\begin{align}
    \mathcal{T}_{AB} =\frac{1}{2} \partial_A \phi \partial_B \phi + \frac{Z}{2} F_{AC}F^C{}_B -   \frac{1}{2} g_{AB} \left[ V + \frac{1}{2} \left(\partial \phi\right)^2 + \frac{Z}{4} F^2 \right]
\end{align}

Our starting point is the $\mathcal{O}\left(\partial^0\right)$ background with a metric ansatz that holds for an arbitrarily boosted frame with arbitrary magnetic field direction, specifically
\begin{align}
    ds^2 &= \left( -g_{tt} u_\mu u_\nu + g_{11} \Pi_{\mu \nu}  + g_{33} b_\mu b_\nu  \right) dx^\mu dx^\nu + g_{rr} d r^2 \, , \\
    \Pi_{\mu \nu} &= \Delta_{\mu \nu}-b_\mu b_\nu \, , \qquad b_\mu=\frac{B_\mu}{|B|} \, . 
\end{align}
where $\{g_{tt},g_{11},g_{33},g_{rr} \}$ are functions of only the holographic direction, and at this point both the four velocity $u^\mu$ and the magnetic field $B^\mu$ are taken to be constant. The background is assumed to be that of a blackhole, namely there exists a horizon at $r_h$ such that
\begin{align}
g_{tt}(r_h) = \frac{1}{g_{rr}(r_h)} = 0   \, .
\end{align}
The temperature $T$ and  \rl{ Hawking's} entropy $s$ are given by 
\begin{align}
    T= \left. \left(  \frac{ g'_{tt}}{4 \pi j}  \right) \right|_{r_h} \, , \qquad s= \left. \frac{e^{-W}}{4} \left( g_{11} \right)^{\frac{3}{2}} \right|_{r_h}\ , 
\end{align}
where we defined
\begin{align}
    j = \sqrt{g_{tt} g_{rr}} \, , \qquad e^{-2 W} = \frac{g_{33}}{g_{11}}
\end{align}
The ansatz for the gauge field is given by 
\begin{align}
    A_\mu = - A_t u_\mu + \mathcal{A}_\mu \, ,  
\end{align}
where $\mathcal{A}_\mu$ is the magnetic potential related to the external magnetic field through
\begin{align}
    B^\mu = \epsilon^{\mu \nu \rho \sigma} u_\nu \partial_\rho \mathcal{A}_\sigma, .  
\end{align}
The $\mathcal{O}\left(\partial^0\right)$ field strength associated to this gauge field is 
\begin{align}
    F = -A'_t u_\mu dr \wedge dx^\mu - \epsilon_{\mu \nu \rho \sigma} u^\rho B^\sigma dx^\mu \wedge dx^\nu \, .
\end{align}
At this order, Maxwell's equations can be integrated once into
\begin{align}
    Q = \frac{\sqrt{-g} Z_V A'_t}{j^2} = \frac{4 s Z_V A'_t}{j}
\end{align}
where $Q$ is an  integration constant. Just as in \cite{Eling:2010hu} we will assume the following horizon behavior
\begin{align}
    A_\mu (r_h) = - \mu u_\mu + \mathcal{A}_\mu \, ,
\end{align} 
where $\mu$ is the associated chemical potential\footnote{As discussed in \cite{Landsteiner:2012kd}, this choice corresponds to introducing the chemical potential in the dual quantum field theory through twisted boundary conditions of the fermions along the thermal circle.}. 
\begin{align}
   \label{E1} \left[E_{AB} \left( \frac{\Pi^{AB}}{g_{11}}-  \frac{2 B^A B^B}{g_{33}} \right) \right]_{r_h} &= 4 \pi T \left[\Delta_2 -\Delta_1 + 2 \pi \lambda T \left(\Delta_1^2-\Delta_2\right)\right] - Z_h \left(4 s e^{W_h} \right)^{-\frac{4}{3} } B^2   \, ,  \\
   \label{E2} \left[ E_{AB} \frac{4 u^A }{j}\left( \delta^B_r- \sqrt{\frac{g_{rr}}{g_{tt}}} u^B \right)\right]_{r_h} &= 4\pi T \left(  \Delta_2 + 2 \Delta_1  \right) - 2 V_h + Z_h \left(4 s e^{W_h} \right)^{-\frac{4}{3} } B^2 +  \frac{Q^2}{16 s^2 Z_h}  \, ,
\end{align}
where the subscript $h$ denotes that the function is evaluated at the horizon, and $\Delta_1$ and $\Delta_2$ are defined as follows
\begin{align}
    \Delta_1 = \left. \frac{1}{j} \frac{g'_{11}}{ g_{11}} \right|_{r_h} \, , \qquad \Delta_2 = \left. \frac{1}{j}\frac{g'_{33}}{g_{33}} \right|_{r_h} \, , 
\end{align}
which from \eqref{E1} and \eqref{E2} can be evaluated as\footnote{There are two solutions for $\{\Delta_1,\Delta_2 \}$, we pick the one that matches the known $B\rightarrow0$ limit.}
\begin{align}
    2 \pi \lambda T \Delta_1 &= \lambda \frac{\left(1 + \Upsilon\right)}{2}  \left(\frac{V_h}{3} - \frac{Q^2}{96 Z_h s^2}  \right)- \frac{Z_h \lambda}{12}  \left(4 s e^{W_h} \right)^{-\frac{4}{3}}B^2  + \frac{1-\Upsilon }{2} \, , \\
    2 \pi \lambda T \Delta_2 &= \lambda \left(2 - \Upsilon \right) \left( \frac{V_h}{3}- \frac{Q^2}{96 Z_h s^2}   \right) - \frac{Z_h \lambda}{3} \left(4 s e^{W_h}\right)^{-\frac{4}{3}} B^2+ \left(\Upsilon- 1 \right) 
\end{align}
with $\Upsilon$ defined as
\begin{align}
    \Upsilon = \sqrt{ 1 + 32 \lambda Z_h^2 s^2 B^2 \left(4 s e^{W_h} \right)^{-\frac{8}{3}} \frac{\lambda \left(4 s e^{W_h} \right)^{\frac{4}{3}} Q^2 + 8 Z_h s^2 \left(B^2 Z_h \lambda - 4 \left(4 s e^{W_h}\right)^{\frac{4}{3}} \left(\lambda V_h - 9 \right)  \right)  }{\left(\lambda Q^2 + 32\left( 3 - \lambda V_h \right) Z_h s^2 \right)^2}  } \, .
\end{align}
Notice that in the absence of magnetic field $\Upsilon=1$. 
Following the standard fluid-gravity procedure we will now assume functions $\{g_{tt}, g_{11},g_{33},g_{rr} \}$ and tensors $\{u_\mu,\Pi_{\mu \nu}, B_\mu \}$ to be slowly varying functions of the boundary coordinates, note that the magnetic field will now satisfy the following Bianchi identity 
\begin{align}\label{BianchiFaraday}
    u^\mu \partial_\mu B^\nu = - \frac{2}{3}\theta B^\nu + \sigma^{\nu \alpha} B_\alpha - \Omega^{\nu \alpha} B_\alpha + u^\nu \partial_\rho B^\rho\, . 
\end{align}
We focus on the near horizon, so it is convenient to switch to Eddington-Finkelstein coordinates using the coordinate transformation
\begin{align}
    u_\mu dx^\mu = u_\nu dy^\mu - \sqrt{\frac{g_{rr}}{g_{tt}} } dv \ , \qquad \Delta_{\mu \nu} dx^\nu = \Delta_{\mu \nu} dy^\nu \, , \qquad dr = dv
\end{align}
In this coordinate system, the metric takes the form\footnote{In principle we should also add higher order corrections to the metric, however at this order in the derivative expansion we expect to be able to chose a fluid frame such that we can reabsorb these corrections into a redefinition of $T$ and $u^\mu$\cite{Eling:2011ms}.}
\begin{align}
    ds^2 = \left(-g_{tt} u_\mu u_\nu + g_{11} \Pi_{\mu \nu} + g_{33} b_\mu b_\nu \right)dy^\mu dy^\nu  + 2 j u_\mu du^\mu dv 
\end{align}
Following \cite{Eling:2011ms} we consider the following projection of Einstein equations evaluated at the horizon
\begin{align}\label{einsteinHorizonForm}
    E_{h} \equiv \left. E_{A B} l^A l^B \right|_{r_h}
\end{align}
where $l^A=(u^\mu,0)$ is a null vector when evaluated at the horizon. We can now evaluate \eqref{einsteinHorizonForm} order by order in gradient expansion, making an extensive use of the Bianchi identity \eqref{BianchiFaraday}, 
\begin{align}
    E_h^{(0)} &= 0 \, , \\
    E_h^{(1)} &= \frac{2 \pi T}{s} \partial_\alpha \left( s u^\alpha \right) \, , \\
    E_h^{(2)} &= - \frac{2}{3}\left[ 1 - \frac{2}{3}\pi T \lambda \left(4\Delta_1 -\Delta_2 \right)  \right] \left[S_{\parallel} - \frac{1}{2} S_{\perp} -  u^\alpha\partial_\alpha W_h \right]^2 -2 \left(1 - 2\pi T \lambda \Delta_1 \right) e^{-2 W_h } \Sigma^2 \\ \nonumber &- \left( 1 - 2 \pi T \lambda \Delta_2 \right)  \sigma_T^2 - \frac{1}{2} \left(\partial_\mu \phi_h\right)^2 - \frac{Z_h}{2} \left(4 s e^W \right)^{-\frac{2}{3} } \left[ e^{2W}  \left(V^\mu \hat B_\mu\right)^2 +  V^\alpha V^\beta \Pi_{\alpha \beta}  \right] \\ \nonumber & + \partial_\alpha\left(s u^\alpha \right) \mathcal{O}\left(\partial\right)
\end{align}
where we defined
\begin{subequations}
    \begin{align}
S_{\perp}  &  = \Pi^{\mu \nu } \partial_{\mu} u_{\nu} ~~ , ~~ S_{\parallel} = \hat{B}^{\mu} \hat{B}^{\nu} \partial_{\mu} u_{\nu} ~~ , ~~ \sigma_T^{\mu \nu } = \hat{\Pi}^{\mu \nu \rho \sigma} \partial_{\rho} u_{\sigma} ~~ , ~~ \Sigma^{\mu} = \Pi^{\mu \sigma } \hat{B}^{\rho} \partial_{(\rho} u_{\sigma)} ~~ , \\ 
\hat{\Pi}_{\mu \nu \rho \sigma}& =     \Pi_{\mu (\rho }     \Pi_{\sigma) \nu }  - \frac{1}{2}  \Pi_{\mu \nu } \Pi_{\rho \sigma}\,,\qquad\qquad\quad
V_\alpha = - \partial_\alpha \mu - \mu a_\alpha \, .
\end{align}
\end{subequations}
We can then rewrite the projection of Einstein's equation at the horizon $E_h=0$ as
\begin{align}
   \label{entropProd} \frac{2 \pi T}{s} \partial_\alpha S^\alpha&= \frac{2}{3}\left[ 1 - \frac{2}{3}\pi T \lambda \left(4\Delta_1 -\Delta_2 \right)  \right] \left[S_{\parallel} - \frac{1}{2} S_{\perp} -  u^\alpha\partial_\alpha W_h \right]^2  \\ \nonumber &+2 \left(1 - 2\pi T \lambda \Delta_1 \right) e^{-2 W_h } \Sigma^2 + \left( 1 - 2 \pi T \lambda \Delta_2 \right)  \sigma_T^2 \\ \nonumber &+ \frac{1}{2} \left(\partial_\mu \phi_h\right)^2 + \frac{Z_h}{2} \left(4 s e^W \right)^{-\frac{2}{3} } \left[ e^{2W}  \left(V^\mu \hat B_\mu\right)^2 +  V^\alpha V^\beta \Pi_{\alpha \beta}  \right] + \mathcal{O}\left(\partial^3 \right) \, , 
\end{align}
where $S^\alpha \equiv s u^\alpha$. Identifying $S^\alpha$ as the entropy current in the dual boundary theory allows us to understand \eqref{entropProd} as the entropy production rate in the system\footnote{ We are finding that the entropy that enters naturally in Eq.~(\ref{entropProd}) obeying positivity of entropy production is Hawking's entropy whereas the entropy of the boundary field theory is expected to be Wald's \cite{Wald:1993nt}. For a discussion on this see the recent paper \cite{Buchel:2024umq}.}. To have a positive definite production rate, in accordance with the second law of thermodynamics, we need the following constraints to hold
\begin{align}
   2 \pi T \lambda \Delta_1 \leq 1 \, , \\
   2 \pi T \lambda \Delta_2 \leq 1 \, , \\
   2 \pi T \lambda \left(4 \Delta_1- \Delta_2 \right) \leq 3  
\end{align}
We can compare this result to the entropy production that follows from the dissipative energy momentum tensor \eqref{eq:stressmagnet} and the U(1) current \eqref{someCurrents}, namely \cite{Hernandez:2017mch}
\begin{align}\label{entropyProdFluid}
   T \partial_\alpha S^\alpha =  \xi_{\parallel} S_{\parallel}^2 + 2 \xi_{\times} S_{\parallel} S_{\perp} + \xi_{\perp} S_{\perp}^2 + 2 \eta_{\parallel} \Sigma^2 + 2 \eta_{\perp} \sigma_T^2 + \sigma_{\parallel} \left(\tilde V^\mu \hat B_\mu \right)^2 + \sigma_\perp \tilde V^\mu \tilde V^\nu \Pi_{\mu \nu} 
\end{align}
where $\tilde V= E_\mu - T \partial_\mu \frac{\mu}{T} $, with $E$ some external electric field. Under first order hydrostatic equations, i.e. $T a_\mu=-  \partial_\mu T$, and in the absence of an electric field one has that $\tilde V^\mu=V^\mu$. Then, comparing \eqref{entropProd} and \eqref{entropyProdFluid} allow us to compute the viscosities and conductivities, finding the shear viscosities
\begin{subequations}\label{eq:shearViscEO}
\begin{align}
 \begin{split}
    \eta_{\parallel} &= \frac{s}{4\pi} e^{-2 W_h}\left(1 - 2\pi T \lambda \Delta_1 \right) \, , 
\end{split}
\\
\begin{split}
  \eta_{\perp} &= \frac{s}{4 \pi} \left(1 - 2 \pi T \lambda \Delta_2 \right) \, ,   
\end{split}   
\end{align}
\end{subequations}
the bulk viscosities
\begin{subequations} \label{eq:BulkViscE0}
\begin{align}
\begin{split}
    \zeta_{\perp } & = \frac{ s  }{4 \pi }   
  \left(  s\frac{\partial  \phi_h }{\partial s }  +\rho \frac{\partial  \phi_h }{\partial \rho  }    +    B \frac{ \partial  \phi_h}{\partial  B }    \right)^2
+  \frac{ s  }{3 \pi } 
   \left(  s\frac{\partial  W_h }{\partial s }  +\rho \frac{\partial  W_h }{\partial \rho  }  + B \frac{\partial W_h}{\partial B} - \frac{1}{2}   \right)^2 \\
   &+ \lambda \frac{sT}{18}\left(1 - 2 s \frac{\partial W_h}{\partial s} - 2 \rho \frac{\partial W_h}{\partial \rho} - 2 B \frac{\partial W_h}{\partial B} \right)^2 \left( 4 \Delta_1 - \Delta_2 \right)  ~~ ,  
\end{split}
     \\ 
 \begin{split}    
     \zeta_{\times} & = \frac{ s  }{4 \pi }  \left(s   \frac{\partial \phi_h}{ \partial s}  +\rho \frac{\partial  \phi_h }{\partial \rho  }    +    B \frac{ \partial  \phi_h}{\partial  B }    \right) \left( s\frac{\partial  \phi_h }{\partial s }  +\rho \frac{\partial  \phi_h }{\partial \rho  }     \right)  \\  & 
     +  \frac{  s }{ 3 \pi } 
 \left( B \frac{\partial W_h}{\partial B}  + s\frac{\partial  W_h }{\partial s }  +\rho \frac{\partial  W_h }{\partial \rho  }- \frac{1}{2}  \right)     \left(  s\frac{\partial  W_h }{\partial s }  +\rho \frac{\partial  W_h }{\partial \rho  }   +1  \right) \\ & + \lambda \frac{sT}{9} \left(1 + s \frac{\partial W}{\partial s} + \rho \frac{\partial W}{\partial \rho} \right) \left(1 - 2 s \frac{\partial W_h}{\partial s}- 2 \rho \frac{\partial W_h}{\partial \rho} - 2 B \frac{\partial W_h}{\partial B} \right)\left( 4 \Delta_1 - \Delta_2 \right) ~~ , 
 \end{split} \\
 \begin{split}
     \zeta_{\parallel}    & =   \frac{  s  }{4 \pi }   \left(  s \frac{ \partial  \phi_h}{\partial s }  + \rho  \frac{  \partial  \phi_h}{\partial \rho }  \right)^2      + \frac{ s  }{3 \pi }\left( s\frac{\partial  W_h }{\partial s }  +\rho \frac{\partial  W_h }{\partial \rho  }  + 1  \right)^2 \\ & - \lambda \frac{2 s T}{9} \left( 1 + s \frac{\partial W_h}{\partial s} + \rho \frac{\partial W_h}{\partial \rho} \right)^2 \left( 4 \Delta_1 - \Delta_2 \right)  ~~ , 
    \end{split}
\end{align}    
\end{subequations}
and the conductivities
\begin{subequations}\label{eq:CondcViscE0}
\begin{align}
        \begin{split}
        \sigma_{\parallel} &= \frac{Z_h}{4 \pi} s e^{2 W_h} \left( 4 s e^{W_h} \right)^{-\frac{2}{3}} \, , 
    \end{split} \\
    \begin{split}
        \sigma_{\perp} &= \frac{Z_h}{4 \pi} s \left( 4 s e^{W_h} \right)^{-\frac{2}{3}} 
    \end{split}
\end{align}
\end{subequations}
when writing equations \eqref{eq:shearViscEO}-\eqref{eq:CondcViscE0} we used that
\begin{align}\label{scalarIdentityPartial}
   u^\mu \partial_\mu F &= \frac{\partial F}{\partial s} u^\mu \partial_\mu s + \frac{\partial F}{\partial \rho} u^\mu \partial_\mu \rho + \frac{\partial F}{\partial B^2} u^\mu \partial_\mu B^2 \\
   &= \nonumber - \left( s \frac{\partial F}{\partial s} + \rho \frac{\partial F}{\partial \rho} \right)\left(S_{\parallel} + S_{\perp} \right) -  B \frac{\partial F}{\partial B} S_{\perp} 
\end{align}
where $F$ here stands for any scalar function of the entropy density, the charge density, and the magnetic field. The identity  \eqref{scalarIdentityPartial} follows from the first order conservation of both the entropy current $s$ and the charge density $\rho$, as well as a projection of the Bianchi identity. These last identities are summarized as follows
\begin{align}
    \partial_\alpha \left(s u^\alpha\right) &= \mathcal{O}\left(\partial^2 \right)\, , \\
    \partial_\alpha \left( \rho u^\alpha \right) &= \mathcal{O}\left(\partial ^2 \right)\, , \\
    u^\alpha \partial_\alpha B^2 &= - 2 S_{\perp} B^2
\end{align}
Note that on the $\lambda \rightarrow 0 $ limit the viscosities match those computed from the BV method. In the isotropic limit, i.e. $B\rightarrow 0$, we recover a single shear and bulk viscosities given by\footnote{We can use relations \eqref{eq:viscosoties9898223321} to find the isotropic viscosities.}
\begin{align}
    \eta &= \frac{s}{4 \pi} \left[1 - \lambda \left( \frac{V_h}{3} - \frac{Q^2}{96 Z_h s^2} \right) \right] \, , \label{etaIsotropic} \\ 
    \zeta &= \frac{s}{4 \pi} \left( \frac{\partial \phi_h}{\partial s} + \frac{\partial \phi_h}{\partial \rho}  \right)^2 \, , 
\end{align}
We note that there are no corrections to the bulk viscosity due to the Gauss-Bonnet term, while the shear viscosity does receive a correction. Note that \eqref{etaIsotropic} is consistent with the bulk viscosity computed in \cite{Brigante:2007nu,Xian2009}. 

\section{Magnetic bulk viscosities in realistic holographic QCD model} 
\label{sec:numericalcheck}

In this section, we explore properties of anisotropic viscosities applied to QCD. Certain aspects of energy-momentum transport in anisotropic states were studied in  holographic QCD literature, see e.g. \cite{Finazzo:2016mhm,Rougemont:2020had,Gursoy:2020kjd} but a comprehensive analysis including both shear and bulk viscosities is missing, which is what we aim to do here. Specifically, we focus on the bulk viscosities in an anisotropic state induced by an external magnetic field\footnote{For anisotropic shear in the same setting see \cite{Demircik:2023lsn}.}. We will employ a ``realistic'' holographic QCD model that goes under the name of V-QCD \cite{Jarvinen:2011qe} (that was build on the earlier improved holographic QCD (IHQCD) model \cite{Gursoy:2007cb,Gursoy:2007er}). This model, which we review below, is realistic in the sense that it not only reproduces the salient features of QCD: confinement, chiral symmetry breaking and asymptotic freedom in the vacuum state, global aspects of the phase diagram of QCD i.e. presence of confinement-deconfinement and chiral symmetry breaking transitions at finite temperature and baryon chemical potential, the baryonic phase, gapped glueball, meson and baryon spectra and so on, but also hosts a range of parameters that makes quantitative agreement with lattice and observational data possible. Our approach involves computing anisotropic viscosities using both the method of background variations and standard fluctuation analysis, aiming at validating the formulae presented in (\ref{eq:viscosoties989821}). This we will do by  numerical evaluation techniques.

The V-QCD model has been used to describe different phases of QCD: the various phenomena at finite-temperature \cite{Alho:2012mh,Arean:2013tja,Jokela:2018ers,Alho:2015zua,Iatrakis:2016ugz}, density 
\cite{Alho:2013hsa,Ishii:2019gta,Jarvinen:2015ofa,Ecker:2019xrw,Hoyos:2020hmq,Demircik:2021zll,Demircik:2020jkc,Hoyos:2021njg,Tootle:2022pvd,CruzRojas:2023ugm,CruzRojas:2024etx}, and magnetic field \cite{Drwenski:2015sha,Demircik:2016nhr,Gursoy:2016ofp,Gursoy:2018ydr,Gursoy:2020kjd}. Transport in the anisotropic phases in this model has been discussed in~\cite{Hoyos:2020hmq,Hoyos:2021njg,CruzRojas:2024etx}. The model is comprised of glue and flavor sectors ($S= S_g + S_f$) with the full back reaction of flavor onto glue sector implemented in the Veneziano limit \cite{Veneziano:1976wm,Veneziano:1979ec}, that is, by taking the number of both colors $N_c$ and flavors $N_f$ to infinity with their ratio $N_f/N_c$ fixed. The glue sector is based on the IHQCD  theory
\cite{Gursoy:2007cb,Gursoy:2007er,Gursoy:2016ebw}
\begin{equation} 
S_g=M_p^3 N_c^2 \int d^{5} x\sqrt{-g}\left[R-\frac{4}{3} g^{M N} \partial_{M} \phi \partial_{N} \phi+V_{g}(\phi)\right] \ .
\end{equation}
where $M_p$ is the five dimensional Planck mass, $R$ is the Ricci scalar, $\phi$ is the dilaton field. The flavor sector takes the tachyonic Dirac-Born-Infeld form~\cite{Bigazzi:2005md,Casero:2007ae,Iatrakis:2010jb,Iatrakis:2010zf}. In this article, we will only study the chirally symmetric phase with zero quark masses in V-QCD, which means that the tachyon vanishes\footnote{Tachyon is dual to the chiral condensate operator in the boundary theory, see \cite{Jarvinen:2021jbd} for a review.}. Moreover, we restrict ourselves to flavor symmetric configurations, i.e., all sources and field components are taken to be equal for different quark flavors. 

In this case the flavor action reduces to
\be\begin{aligned}
S_{f}&= -  \frac{1}{2}M_p^3 N_c N_f \int d^5x\,V_{f0}(\phi) 
\sqrt{-\det (g_{MN} + w(\phi) F_{MN})} 
\ ,
\label{generalact2}
\end{aligned}\ee
where $F_{MN}$ is the field strength tensor for the Abelian gauge field. We consider chirally symmetric deconfined configuration at finite temperature and magnetic field but at zero density. Magnetic field is introduced by turning on the gauge potential $A_M= \{0,-yB/22,xB/2,0,0\}$. 
The background geometry can be taken to be of the form~\eqref{eq:metricansatz0}, and a nonzero dilaton depending on the radial coordinate also needs to be included.

The V-QCD potentials $V_g$, $V_f$ and $w$
are  constructed with a set of parameters tuned to experimental data, lattice QCD, and predictions from perturbative QCD. In this work we use two different set of potentials. First, we use the 7a potential set \cite{Jokela:2018ers,Ishii:2019gta,Jokela:2020piw} which was fitted among other things to lattice data for the QCD equation of state at finite temperature and small density, setting $N_f/N_c=1$.
Second, we use another potential set that was employed in \cite{Drwenski:2015sha,Demircik:2016nhr} where the flavor-color ratio was chosen as $N_f/N_c = 0.1$, so that the effects of flavor to the geometry are reduced, to compare with the 7a results and to study which features of viscosities that remain robust regardless of choice of potentials. This latter set was not quantitatively fitted to QCD data, but chosen to agree with various qualitative features.
For more details on V-QCD, we refer to \cite{Jarvinen:2011qe} and \cite{Jarvinen:2021jbd}.

\begin{figure*}
     \includegraphics[height=0.045\textwidth]{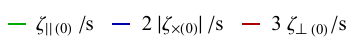}
    \qquad\qquad\qquad\qquad
    \includegraphics[height=0.045\textwidth]{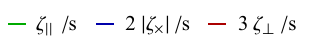}
    \includegraphics[height=0.31\textwidth]{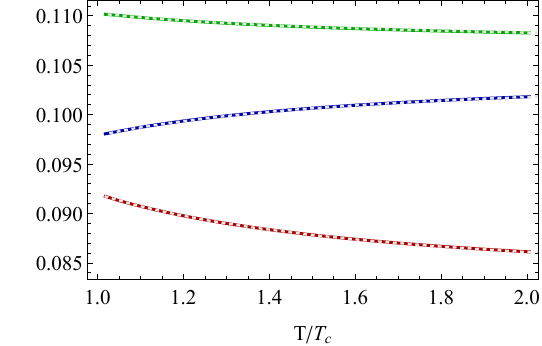}
    \includegraphics[height=0.31\textwidth]{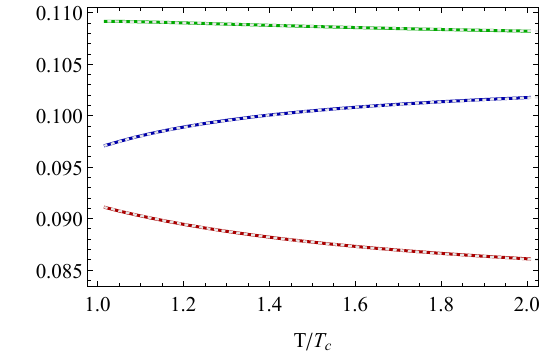}

    \includegraphics[height=0.05\textwidth]{legc.pdf}\\
    \includegraphics[height=0.32\textwidth]{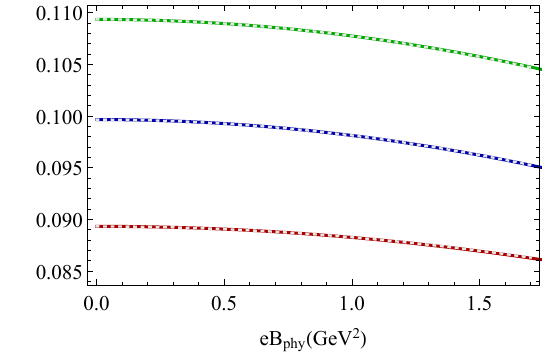}    
    \caption{\small Top-Left: Temperature dependence  anisotropic bulk viscosities at  $B\rightarrow0$ limit. Top-Right: Temperature dependence  anisotropic bulk viscosities at  $eB_{phy}=0.5GeV^2$.
    Bottom: Magnetic field  dependence  anisotropic bulk viscosities at  $T/T_c=1.25$. Colored solid curves are obtained from
standard fluctuation analysis while dotted gray curves are from background variations. In both figures, the bulk viscosities are scaled differently (since there is ordering of $\zeta_{\parallel}/s>|\zeta_{\times}|/s>\zeta_{\perp}/s$) for better visibility.}
    \label{p1}
\end{figure*}
\begin{figure*}
    \includegraphics[height=0.32\textwidth]{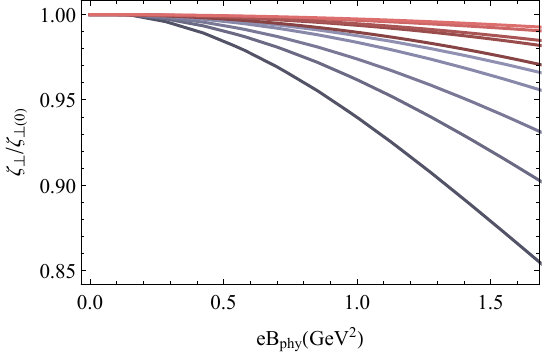}
    \includegraphics[height=0.32\textwidth]{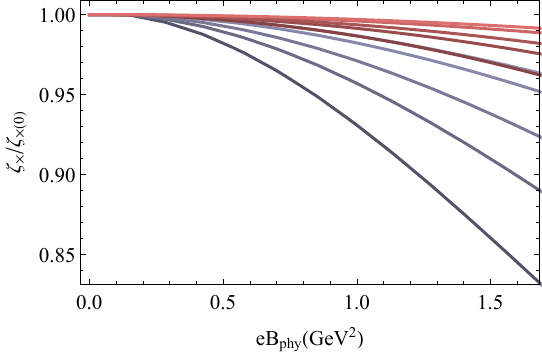}\\
\vspace{1cm}
 \includegraphics[height=0.32\textwidth]{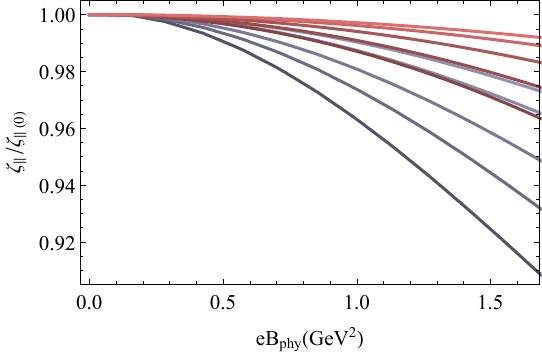}    
    \caption{\small Comparison of different V-QCD potentials: Red and gray curves denote the magnetic field dependence of anisotopic bulk viscosities (at fixed temperature) for the potentials used in \cite{Jokela:2018ers,Ishii:2019gta,Jokela:2020piw} and \cite{Drwenski:2015sha,Demircik:2016nhr}   respectively. For the both set of the curves, the darkness decreases with temperature. The considered temperature values are $T/T_c=\{1.29, 1.42, 1.54, 1.72, 1.84\}$. The bulk viscosities are scaled with $B\rightarrow0$ values i.e. $\zeta_{\perp(0)}$, $\zeta_{\times(0)}$, $\zeta_{||(0)}$.   }
    \label{p2}
\end{figure*}

One first numerically constructs the the black hole solutions in the V-QCD model with the desired ranges of magnetic field and temperature. We follow~\cite{Alho:2013hsa} and use the scale function $A$ in~\eqref{eq:metricansatz} as the holographic radial coordinate for the numerical solutions.

The next step in the standard fluctuation analysis (GPR) is to solve the fluctuation equations (\ref{fluceqsd}) with the UV boundary conditions given in (\ref{eq:boundaryconditions}). As mentioned earlier, it is sufficient to solve them for vanishing frequency. Therefore, we set $\omega=0$ and solve the fluctuation equations from horizon to boundary  and construct a linear combinations of these solutions that satisfy the UV boundary conditions (\ref{eq:boundaryconditions}). From the set of the bulk solutions constructed in this procedure, the boundary coefficients $\Sigma^I$ and $\Delta^J$ can be read off and consequently $\mathcal{F}_{IJ}$ can be constructed. Finally, the magnetic bulk viscosities can be computed by plugging $\mathcal{F}_{IJ}$ into the equations (\ref{eq:zetformulae}). 

Using the BV method, it is possible to directly compute the bulk viscosities 
after the construction of the background. One 
needs  to read off the horizon values $\{\phi_h, A_h, W_h\}$, computing the entropy density $s$ and constructing them as functions of $\{T, B\}$. Then, the computation of the derivatives yields the viscosities as provided in  (\ref{eq:viscosoties989821}). 

The results illustrating the temperature and magnetic field dependence of anisotropic bulk viscosities are presented in Figure~\ref{p1}. 
The top left and right plots  depict temperature dependence of bulk viscosities in the $B\rightarrow0$ limit and at $eB_{\text{phy}}=0.5\text{GeV}^2$, respectively, where $e$ denotes the elementary electric charge. The bottom plot shows the magnetic field dependence of bulk viscosities at a fixed temperature of $T/T_c=1.25$, where $T_c = 120$~MeV is the critical temperature for deconfinement phase transition for 7a at $B=0$. In all three plots, green, blue, and red curves correspond to $\zeta_{\perp}/s$, $|\zeta_{\times}|/s$, and $\zeta_{\parallel}/s$, respectively, scaled by numerical factors as indicated in the plot legends in order to make the details better visible. Solid colored curves are obtained through standard fluctuation analysis (GPR), while the dotted gray are obtained through the formulae (\ref{eq:viscosoties989821}) derived using the background variation method. We observe that the two methods, GPR and BV are in perfect agreement. 
We observe that the magnitudes of bulk viscosity to entropy ratios are ordered as $\zeta_{\parallel}/s>|\zeta_{\times}|/s>\zeta_{\perp}/s$. These inequalities are not obvious from the analytic formulae (\ref{eq:viscosoties989821}). 
Moreover, the ratio of $\zeta_{\parallel}/s$ is larger than the universal value ($1/4\pi$) of the shear viscosity-entropy ratio, whereas the ratios of $\zeta_{\times}/s$ and $\zeta_{\perp}/s$ are smaller, but still of the same order. 
As evident from the top plots in Figure~\ref{p1},  $\zeta_\parallel/s$ and $\zeta_\perp/s$ decrease with increasing temperature while $|\zeta_\times|/s$ exhibits the opposite behaviour. When the magnetic field is increased, they all decrease as seen from the bottom plot in Figure~\ref{p1}. We also note that both the magnetic field and temperature dependence of bulk viscosities are overall very mild. This implies that the error one would make by treating them constant would be small in practice.

In Figure~\ref{p2}, we compare these results with a different set of V-QCD potential sets used in \cite{Drwenski:2015sha,Demircik:2016nhr}. Bulk viscosities, normalized to their $B\rightarrow0$ values ($\zeta_{\perp}/\zeta_{\perp(0)}$,  $\zeta_{\times}/\zeta_{\times(0)}$,   $\zeta_{\parallel}/\zeta_{\parallel(0)}$), are depicted in separate plots. Each plot includes two sets of curves showing the magnetic field dependence at fixed temperature: red curves for the 7a potentials and gray curves for the potentials used in \cite{Drwenski:2015sha,Demircik:2016nhr}. Temperature values considered for each potential set are identical, i.e., $T/T_c=\{1.29, 1.42, 1.54, 1.72, 1.84\}$. For each set, darkness of the curves decreases with increasing temperature.

Even though the two potential choices are rather different, the results have obvious similarities.
First, the magnetic field dependence becomes more pronounced at lower temperatures, suggesting that neglecting the magnetic field dependence of bulk viscosities in simulations would be a better approximation at higher temperatures. Second, all three ratios, $\zeta_{\perp}/\zeta_{\perp(0)}$, $\zeta_{\times}/\zeta_{\times(0)}$, $\zeta_{\parallel}/\zeta_{\parallel(0)}$, decrease with increasing magnetic field.  This decrease is particularly notable at lower temperatures because
of more pronounced magnetic field dependence as mentioned earlier. Consequently, dissipative processes governed by these transport coefficients become less significant as temperature decreases and magnetic field strengthens, hence the plasma approaches the ideal fluid limit. 

To conclude this section, we discuss consistency checks of generic hydrodynamic relations (see \cite{Hernandez:2017mch}) that 
we carried out numerically: 
i) the bulk viscosities satisfy the Onsager relations 
in the entire magnetic field range, ii) They obey the anticipated  relations at $B=0$, i.e. Eq.~\eqref{eq:viscosoties9898223321}., iii) The 
 combination (from Eq.~\eqref{eq:viscosoties9898223321}) corresponding to $\zeta$ at $B\rightarrow0$ limit, exhibits a temperature dependence that aligns with the earlier observations \cite{Buchel:2011wx}.

\section{Discussion}\label{sec:disc}

In this work we studied transport in anisotropic phases of matter by means of holography, focusing particularly on anisotropy induced by a background magnetic field. Firstly, we established a connection between the GPR method for computing bulk viscosity, which relies on solving fluctuation equations, and the EO method, which only requires varying the background. Unlike the holographic formula for electric conductivity or shear viscosity, see e.g. \cite{Iqbal:2008by}, this connection is non-trivial because on the GPR side it involves a numerical coefficient that represents the flow of gravitational fluctuations from the UV boundary to the horizon. We showed that the GPR formula for bulk viscosity can be written as the EO formula by absorbing fluctuations  into background variations and turning off the UV boundary condition for the fluctuations using reparametrization invariance. This connection was then used to compute a formula for the three distinct bulk viscosities that occur in an anisotropic phase induced by an external magnetic field, which is solely given in terms of background field variations. From the expression in terms of variations of the background fields, it follows immediately that the constraints on the anisotropic bulk viscosities coming from the Second Law of Thermodynamics are satisfied. Furthermore, we validated the expressions and also the connections between the background variation method and GPR method through a numerical comparison of the results obtained from both approaches. V-QCD, a realistic bottom-up effective holographic model of QCD, was used for this purpose. 
In addition, the numerical results present a thorough investigation of magnetic bulk viscosities, including their dependencies on temperature, magnetic field, and model parameters.

In addition to anisotropic bulk viscosities, we also derived horizon formulae for anisotropic conductivities induced by a background magnetic field. It is found that the ratio of the longitudinal and transverse conductivity can be universally related to the ratio of the magnetic shear viscosities. This relation will however be modified by higher derivative corrections to the gravity theory as these modify shear viscosity but not conductivity, see section \ref{sec:EO}. 

The fact that transport coefficients of the boundary field theory can solely be expressed in terms of horizon data is consistent with the generic lore of the membrane paradigm \cite{DamourThese, Damour:2008ji,Price:1986yy}, see also \cite{Iqbal:2008by,Eling:2009sj}. Expressing fluctuations of background fields in terms of background variations provide this connection between the boundary and the horizon. Apart from the practical advantage of using background variations --- one needs to solve differential equations only once for the background, not a second time also for the fluctuations as in the GPR method --- it also reveals an alternative formulation of the fluid-gravity correspondence \cite{Bhattacharyya:2007vjd}. This is clear from the fact that background variations connect the standard holographic calculation of retarded Green's functions to the method of Eling and Oz \cite{Eling:2009sj} which is essentially based on recasting the null-focusing equation at the horizon as the entropy balance law (local second law). EO, in turn, is based on the fluid-gravity correspondence. 

However, we emphasize that, while the EO approach is blind to dissipationless transport, the method of background variations that we introduce here can in principle also provide analytic formulae for transport that do not generate entropy. \rl{ It will be interesting to employ our method to study anomalous transport. More specifically, anomalies caused by dynamical gauge fields  (e.g. gluons in QCD) which are expected receive radiative
corrections. Even though this has been addressed in holography in \cite{Klebanov:2002gr,Gursoy:2014ela,Jimenez-Alba:2014iia,Gallegos:2018ozs} --- see \cite{Gallegos:2024qxo} for the latest study ---   
general analytic expressions that relate transport to horizon data are still missing. Since the BV method is shown to be successful for massive fluctuations and is not built on
positive entropy production, it can potentially be used to investigate contributions of   dynamical gauge fields to anomalous transport. In addition, the BV method can perhaps be employed to study the Hall viscosity \cite{Hoyos2019}, \cite{Hernandez:2017mch} in three dimensions which is another non-dissipative transport coefficient that can appear in hydrodynamics subject to a background magnetic field. Lastly, although not non-dissipative, it is found for shear viscosity that metric fluctuations become massive in the case of translation symmetry breaking \cite{Hartnoll_2016,Alberte_2016}. It might be that the method can be of use here as well. 
} 

Another question is whether it is possible to incorporate higher derivative corrections in our method of background variations. In principle there is no obstacle apart from the technicalities. In fact, that we were able to extend the Eling-Oz method to include these corrections in section \ref{sec:EO} strongly suggests that this should be possible. We plan to turn to this problem in future work.

We close with the hope that the method we introduced in this paper, and its companion \cite{Demircik:2023lsn}, will prove useful in the vast field of applications of gauge-gravity duality to both particle and condensed-matter physics.

\section{Acknowledgments}

We are grateful to Alex Buchel, Richard Davison, Niko Jokela, Govert Nijs and Francisco Peña-Benitez for discussions. This work was supported, in part by the Netherlands Organisation for Scientific Research (NWO) under the VICI grant VI.C.202.104. T.D. 
acknowledges the support of the Narodowe Centrum Nauki (NCN) Sonata Bis Grant No.
2019/34/E/ST3/00405. M.~J. has been supported by an appointment to the JRG Program at the APCTP through the Science and Technology Promotion Fund and Lottery Fund of the Korean Government. M.~J. has also been supported by the Korean Local Governments -- Gyeong\-sang\-buk-do Province and Pohang City -- and by the National Research Foundation of Korea (NRF) funded by the Korean government (MSIT) (grant number 2021R1A2C1010834).

\end{document}